\documentclass[superscriptaddress,twocolumn,floatfix]{revtex4-2}
\usepackage{graphicx}
\usepackage{bbm}
\usepackage{physics}
\usepackage{mathtools}
\usepackage{amssymb,amsmath,amsfonts,dsfont}
\usepackage{color}
\usepackage{adjustbox}
\usepackage{tabularx}
\usepackage{array}
\usepackage{multirow}
\usepackage{tabu}
\usepackage{tikz}
\usepackage{makecell}
\usepackage{url}
\usepackage{booktabs}

\newcommand{\PP}[1][]{
  \ifthenelse{\isempty{#1}}
    {\mathbbm{P}}
    {\mathbbm{P}\left[#1\right]}
}
\newcommand{\EE}[1][]{
  \ifthenelse{\isempty{#1}}
    {\mathbbm{E}}
    {\mathbbm{E}\left[#1\right]}
}

\begin{document}
\title{Quantum convolutional neural network for classical data classification}
\author{Tak Hur}
\thanks{These authors contributed equally to this work and are listed in alphabetical order.}
\affiliation{Department of Physics, Imperial College London, London, SW7 2BW, UK}
\author{Leeseok Kim}
\thanks{These authors contributed equally to this work and are listed in alphabetical order.}
\affiliation{Department of Electrical and Computer Engineering, University of New Mexico, Albuquerque, NM 87131, USA}
\author{Daniel K. Park}
\email{dkp.quantum@gmail.com}
\affiliation{Sungkyunkwan University Advanced Institute of Nanotechnology, Suwon, 16419, Republic of Korea}

\begin{abstract}
With the rapid advance of quantum machine learning, several proposals for the quantum-analogue of convolutional neural network (CNN) have emerged. In this work, we benchmark fully parameterized quantum convolutional neural networks (QCNNs) for classical data classification. In particular, we propose a quantum neural network model inspired by CNN that only uses two-qubit interactions throughout the entire algorithm. We investigate the performance of various QCNN models differentiated by structures of parameterized quantum circuits, quantum data encoding methods, classical data pre-processing methods, cost functions and optimizers on MNIST and Fashion MNIST datasets. In most instances, QCNN achieved excellent classification accuracy despite having a small number of free parameters. The QCNN models performed noticeably better than CNN models under the similar training conditions. Since the QCNN algorithm presented in this work utilizes fully parameterized and shallow-depth quantum circuits, it is suitable for Noisy Intermediate-Scale Quantum (NISQ) devices.
\keywords{Quantum machine learning \and Convolutional neural network \and Deep learning}
\end{abstract}

\maketitle
\def\one{{\mathchoice {\rm 1\mskip-4mu l} {\rm 1\mskip-4mu l} {\rm \mskip-4.5mu l} {\rm 1\mskip-5mu l}}}
\section{Introduction}
\label{sec:intro}
Machine learning techniques with artificial neural networks are ubiquitous in modern society as the ability to make reliable predictions from the vast amount of data is essential in various domains of science and technology. A convolutional neural network (CNN) is one such example, especially for data with a large number of features. It effectively captures spatial correlation within data and learns important features~\cite{726791}, which is shown to be useful for many pattern recognition problems, such as image classification, signal processing, and natural language processing~\cite{lecun_deep_2015}. It has also opened the path to Generative Adversarial Networks (GANs)~\cite{10.5555/2969033.2969125}. CNNs are also rising as a useful tool for scientific research, such as in high energy physics~\cite{Aurisano_2016,Acciarri_2017}, gravitational wave detection~\cite{GEORGE201864} and statistical physics~\cite{doi:10.7566/JPSJ.86.063001}. By all means, the computational power required for the success of machine learning algorithms increases with the volume of data, which is increasing at an overwhelming rate. With the potential of quantum computers to outperform any foreseeable classical computers for solving certain computational tasks, Quantum machine learning (QML) has emerged as the potential solution to address the challenge of handling an ever-increasing amount of data. For example, several innovative quantum machine learning algorithms have been proposed to offer speedups over their classical counterparts~\cite{lloyd2013quantum,qPCA,10.5555/2871393.2871400,PhysRevLett.113.130503_QSVM,NEURIPS2019_16026d60,blank2020quantum}. Motivated by the benefits of CNN and the potential power of QML, Quantum Convolutional Neural Network (QCNN) algorithms have been developed~\cite{cong_quantum_2019,kerenidis2019quantum,liu2019hybrid,henderson_quanvolutional_2020,chen2020quantum,li_quantum_2020,maccormack_branching_2020,wei_quantum_2021,mangini_quantum_2021} (see Appendix~\ref{sec:appA} for a brief summary and comparison of other approaches to QCNN). Previous constructions of QCNN have reported success in developing efficient quantum arithmetic operations that exactly implement the basic functionalities of classical CNN or in developing parameterized quantum circuits inspired by key characteristics of CNN. While the former likely requires fault-tolerant quantum devices, the latter has been focused on quantum data classification. In particular, Cong et al. proposed a fully parameterized quantum circuit (PQC) architecture inspired by CNN and demonstrated its success for some quantum many-body problems~\cite{cong_quantum_2019}. However, the study of fully parameterized QCNN for performing pattern recognition, such as classification, on classical data is missing.

In this work, we present a fully parameterized quantum circuit model for QCNN that solves supervised classification problems on classical data. In a similar vein to~\cite{cong_quantum_2019}, our model only uses two-qubit interactions throughout the entire algorithm in a systematic way. The PQC models---also known as variational quantum circuits~\cite{cerezo2020variational}---are attractive since they are expected to be suitable for Noisy Intermediate-Scale Quantum (NISQ) hardware~\cite{Preskill2018quantumcomputingin,bharti2021noisy}. Another advantage of QCNN models for NISQ computing is their intrinsically shallow circuit depth. Furthermore, QCNN models studied in this work exploit entanglement, which is a global property, and hence have the potential to transcend classical CNN that is only able to capture local correlations. We benchmark the performance of the parameterized QCNN with respect to several variables, such as quantum data encoding methods, structures of parameterized quantum circuits, cost functions, and optimizers using two standard datasets, namely MNIST and Fashion MNIST, on Pennylane~\cite{bergholm2020pennylane}. The quantum encoding benchmark also examines classical dimensionality reduction methods, which is essential for early quantum computers with a limited number of logical qubits. The various QCNN models tested in this work employs a small number of free parameters, ranging from 12 to 51. Nevertheless, all QCNN models produced high classification accuracy, with the best case being about $99\%$ for MNIST and about $94\%$ for Fashion MNIST. Moreover, we discuss a QCNN model that only requires nearest neighbour qubit interactions, which is a desirable feature for NISQ computing. Comparing classification performances of QCNN and CNN models shows that QCNN is more favorable than CNN under the similar training conditions for both benchmarking datasets.

The remainder of the paper is organized as follows. Section~\ref{sec:theoretical} sets the theoretical framework of this work by describing the classification problem, the QCNN algorithm, and various methods for encoding classical data as a quantum state. Section~\ref{sec:bmv} describes variables of the QCNN model, such as parameterized quantum circuits, cost functions, and classical data pre-processing methods, that are subject to our benchmarking study. Section~\ref{sec:simulation} compares and presents the performance of various designs of QCNN for binary classification of MNIST and Fashion MNIST datasets. Conclusions are drawn and directions for future work are suggested in Section~\ref{sec:conclusion}.

\section{Theoretical framework}
\label{sec:theoretical}
\subsection{Classification}
Classification is an example of pattern recognition, which is a fundamental problem in data science that can be effectively addressed via machine learning. The goal of $L$-class classification is to infer the class label of an unseen data point $\tilde{x} \in \mathbbm{C}^N$, given a labelled data set $$\mathcal{D} = \left\{ (x_1, y_1), \ldots, (x_M, y_M) \right\} \subset \mathbbm{C}^N\times\{0,1,\ldots,L-1\}.$$
The classification problem can be solved by training a parameterized quantum circuit. Hereinafter, we refer to fully parameterized quantum circuits trained for machine learning tasks as Quantum Neural Network (QNN). For this supervised classification task, a QNN is trained by optimizing the parameters of quantum gates so as to minimize the cost function 
$$ C(\boldsymbol{\theta}) = \sum_{i=1}^{M} \alpha_i c(y_i,f(x_i,\boldsymbol{\theta}))$$
where $f(x_i,\boldsymbol{\theta})$ is the machine learning model defined by the set of parameters $\boldsymbol{\theta}$ that predicts the label of $x_i$, $c(a,b)$ quantifies the dissimilarity between $a$ and $b$, and $\alpha_i$ is a weight that satisfies $\sum_{i=1}^{M}\alpha_i=1$. After the training is finished, the class label for the unseen data point $\tilde{x}$ is determined as $ \tilde{y} = f(\tilde{x},\boldsymbol{\theta}^{*}),$ where $\boldsymbol{\theta}^{*} = \arg\min_{\boldsymbol{\theta}} C(\boldsymbol{\theta})$. If the problem is restricted to binary classification (i.e. $L=2$), the class label can be inferred from a single-qubit von Neumann measurement. For example, the sign of an expectation value of $\sigma_z$ observable can represent the binary label~\cite{park2021robust}. Hereinafter, we focus on the binary classification, albeit potential future work towards multi-class classification will be discussed in Sec.~\ref{sec:conclusion}.

\subsection{Quantum Convolutional Neural Network}
\label{sec:qcnn}
An interesting family of quantum neural networks utilizes tree-like (or hierarchical) structures~\cite{grant_hierarchical_2018} with which the number of qubits from a preceding layer is reduced by a factor of two for the subsequent layer. Such architectures consist of $O(\log(n))$ layers for $n$ input qubits, thereby permitting shallow circuit depth. Moreover, they can avoid one of the most critical problems in the PQC based algorithms known as ``barren plateau", thereby guaranteeing the trainability~\cite{pesah2020absence}. These structures also make a natural connection to the tensor network, which serves as a useful ground for exploring many-body physics, neural networks, and the interplay between them.

\begin{figure*}[ht]
    \centering
    \includegraphics[width=0.8\textwidth]{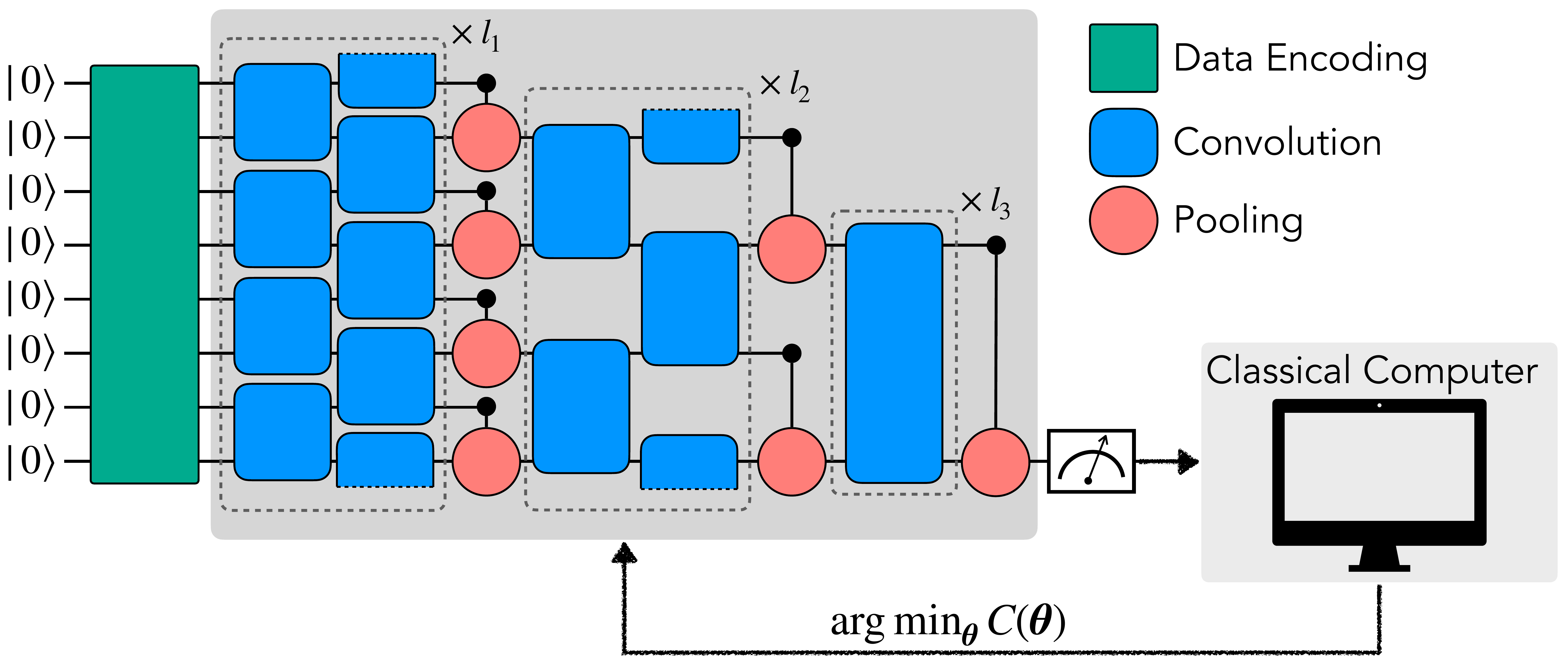}
    \caption{A schematic of the QCNN algorithm for an example of 8 input qubits. The quantum circuit consists of three parts: quantum data encoding (green rectangle), convolutional filters (blue rounded rectangle), and pooling (red circle). The quantum data encoding is fixed in a given structure of QCNN, while the convolutional filter and pooling use parameterized quantum gates. There are three layers in this example, and in each layer, multiple convolutional filters can be applied. The number of filters for $i$th layer is denoted by $l_i$. In each layer, the convolutional filter applies the same two-qubit ansatz to nearest neighbour qubits in a translationally-invariant way. Similarly, pooling operations within the layer use the same ansatz. In this example, the pooling operation is represented as a controlled unitary transformation, which is activated when the control qubit is $1$. However, general controlled operations can also be considered. The measurement outcome of the quantum circuit is used to calculate the user-defined cost function. The classical computer is used to compute the new set of parameters based on the gradient, and the quantum circuit parameters are updated accordingly for the subsequent round.}
    \label{fig:1}
\end{figure*}

The progressive reduction of the number of qubits is analogous to the pooling operation in CNN. A distinct feature of the QCNN architecture is the translational invariance, which forces the blocks of parameterized quantum gates to be identical within a layer. The quantum state resulting from an $i$th layer of QCNN can be expressed as
\begin{equation}
    \ketbra{\psi_{i}(\boldsymbol{\theta}_i)}{\psi_{i}(\boldsymbol{\theta}_i)} = \Tr_{B_i}(U_i(\boldsymbol{\theta}_i)\ketbra{\psi_{i-1}}{\psi_{i-1}}U_i(\boldsymbol{\theta}_i)^{\dagger}),
\end{equation}
where $\Tr_{B_i}(\cdot)$ is the partial trace operation over subsystem $B_i\in \mathbb{C}^{2^{n/2^{i}}}$, $U_i$ is the parameterized unitary gate operation that includes quantum convolution and the gate part of pooling, and $|\psi_0\rangle = |0\rangle^{\otimes n}$. Following the existing nomenclature, we refer to the structure (or template) of the parameterized quantum circuit as \textit{ansatz}. In our QCNN architecture, $U_i$ always consists of two-qubit quantum circuit blocks, and the convolution and pooling part each uses the identical quantum circuit blocks within the given layer. Since a two-qubit gate requires 15 parameters at most~\cite{PhysRevA.69.032315}, in $i$th layer consisting of $l_i>0$ independent convolutional filter and one pooling operation the maximum number of parameters subject to optimization is $15(l_i+1)$. Then the total number of parameters is at most $15\sum_{i=1}^{\log_2(n)}(l_i+1)$ if the convolution and pooling operations are iterated until only one qubit remains. One can also consider an interesting hybrid architecture in which the QCNN layers are stacked until $m$ qubits are remaining and then a classical neural network takes over from the $m$ qubit measurement outcomes. In this case, the number of quantum circuit parameters is less than the maximum number given above. Usually, $l_i$ is set to be a constant. Therefore, the number of parameters subject to optimization grows as $O(\log(n))$, which is an exponential reduction compared to the general hierarchical structure discussed in Ref.~\cite{grant_hierarchical_2018}. This also implies that the number of parameters can be suppressed double-exponentially with the size of classical data if the exponentially large state space is fully utilized for encoding the classical data. An example quantum circuit for a QCNN algorithm with eight qubits for binary classification is depicted in Fig.~\ref{fig:1}. 
Generalizing Fig.~\ref{fig:1} to larger systems can be done simply by connecting all neighboring qubits with the two-qubit parameterized gates in the translationally invariant way.

The optimization of the gate parameters can be carried out by iteratively updating the parameters based on the gradient of the cost function until some condition for the termination is reached. The cost function gradient can be calculated classically or by using a quantum computer via \textit{parameter-shift rule}~\cite{PhysRevLett.118.150503,PhysRevA.98.032309,PhysRevA.99.032331}. When the parameter-shift rule is used, QCNN requires an exponentially smaller number of quantum circuit executions compared to the general hierarchical structures inspired by tensor networks (e.g. tree tensor network) in Ref.~\cite{grant_hierarchical_2018}. While the latter uses $O(n)$ runs, the former only uses $O(\log(n))$ runs.

\subsection{Quantum data encoding}

Many machine learning techniques transform input data $\mathcal{X}$ into a different space to make it easier to work with. This transformation $\phi: \mathcal{X} \rightarrow \mathcal{X}'$ is often called a feature map. In quantum computing, the same analogy can be applied to perform a \textit{quantum feature map}, which acts as $\phi: \mathcal{X} \rightarrow \mathcal{H}$ where the vector space $\mathcal{H}$ is a Hilbert space~\cite{PhysRevLett.122.040504}. In fact, such feature mapping is mandatory when one uses quantum machine learning on classical data, since classical data must be encoded as a quantum state~\cite{PhysRevLett.113.130503_QSVM,qPCA,PhysRevLett.100.160501,ffqram,9259210,araujo_divide-and-conquer_2021}. The quantum feature map $x \in \mathcal{X} \rightarrow \ket{\phi(x)} \in \mathcal{H}$ is equivalent to applying a unitary transformation $U_{\phi}(x)$ to the initial state $\ket{0}^{\otimes{n}}$ to produce $U_{\phi}(x)\ket{0}^{\otimes{n}} = \ket{\phi(x)}$, where $n$ is the number of qubits. This refers to the green rectangle in Fig.~\ref{fig:1}.

There exist numerous structures of $U_{\phi}(x)$ to encode the classical input data $x$ into a quantum state.
In this work, we benchmark the performance of the QCNN algorithm with several different quantum data encoding techniques. These techniques are explained in detail in this section.

\subsubsection{Amplitude encoding}
One of the most general approaches to encode classical data as a quantum state is to associate normalized input data with probability amplitudes of a quantum state. This encoding scheme is known as the amplitude encoding (AE). The amplitude encoding represents input data of $x = (x_1, ..., x_{N})^T$ of dimension $N = 2^n$ as amplitudes of an $n$-qubit quantum state $\ket{\phi(x)}$ as
\begin{equation}
\label{eq:amplitude}
    U_{\phi}(x) : x \in \mathbb{R}^N \rightarrow \ket{\phi(x)} = \frac{1}{\|x\|} \sum_{i=1}^{N} x_i \ket{i},
\end{equation}
where $\ket{i}$ is the $i$th computational basis state. Clearly, with amplitude encoding, a quantum computer can represent exponentially many classical data. This can be of great advantage in QCNN algorithms. Since the number of parameters subject to optimization scales as $O(\log(n))$ (see Sec.~\ref{sec:qcnn}), the amplitude encoding reduces the number of parameters doubly-exponentially with the size (i.e. dimension) of the classical data. However, the quantum circuit depth for amplitude encoding usually grows as $O(poly(N))$, although there exists a method to reduce it to $O(\log(N))$ at the cost of increasing the number of qubits to $O(N)$~\cite{araujo_divide-and-conquer_2021}.

\subsubsection{Qubit encoding}

The computational overhead of amplitude encoding motivates qubit encoding, which uses a constant quantum circuit depth while using $O(N)$ number of qubits. The qubit encoding embeds one classical data point $x_i$, that is rescaled to lie between $0$ and $\pi$, into a single qubit as $\ket{\phi(x_i)} = \cos(\frac{x_i}{2})\ket{0} + \sin(\frac{x_i}{2})\ket{1}$ for $i = 1, ..., N$. Hence, the qubit encoding maps input data of $x = (x_1,\ldots,x_N)^T$ to $N$ qubits as
\begin{align}
\label{eq:qubit}
    U_{\phi}(x) : x \in \mathbb{R}^N  \rightarrow & \ket{\phi(x)} \nonumber \\ & = \bigotimes_{i=1}^{N} (\cos(\frac{x_i}{2})\ket{0} + \sin(\frac{x_i}{2})\ket{1}),
\end{align}
where $x_i\in \lbrack 0,\pi)$ for all $i$. The encoding circuit can be expressed with a unitary operator $U_{\phi}(x) = \bigotimes_{j=1}^{N} U_{x_j}$ where 
\begin{equation*}
    U_{x_j} = e^{-i \frac{x_j}{2} \sigma_y} := 
    \begin{bmatrix}
    \cos(\frac{x_j}{2}) & -\sin(\frac{x_j}{2}) \\
    \sin(\frac{x_j}{2}) & \cos(\frac{x_j}{2})
    \end{bmatrix}.
\end{equation*}

\subsubsection{Dense qubit encoding}

In principle, since a quantum state of one qubit can be described with two real-valued parameters, two classical data points can be encoded in one qubit. Thus the qubit encoding described above can be generalized to encode two classical vectors per qubit by using rotations around two orthogonal axes~\cite{PhysRevA.102.032420}. By choosing them to be the $x$ and $y$ axes of the Bloch sphere, this method, which we refer to as dense qubit encoding, encodes $x_j = (x_{j_1}, x_{j_2})$ into a single qubit as 
$$\ket{\phi(x_j)} = e^{-i\frac{x_{j_2}}{2}\sigma_{y}}e^{-i\frac{x_{j_1}}{2}\sigma_{x}} \ket{0}.$$
Hence, the dense qubit encoding maps an $N$-dimensional input data $x = (x_1,\ldots,x_N)^T$ to $N/2$ qubits as
\begin{align}
\label{eq:dense}
    U_{\phi}(x) : x \in \mathbb{R}^{N} \rightarrow & \ket{\phi(x)} \nonumber \\
    & = \bigotimes_{j=1}^{N/2} \left( e^{-i\frac{x_{N/2+j}}{2}\sigma_{y}}e^{-i\frac{x_{j}}{2}\sigma_{x}} \ket{0} \right).
\end{align}
Note that there is freedom to choose which pair of classical data to be encoded in one qubit. In this work, we chose the pairing as shown in Eq.~(\ref{eq:dense}), but one may choose to encode $x_{2j-1}$ and $x_{2j}$ in $j$th qubit.


\subsubsection{Hybrid Encoding}
\label{sec:hybrid_enc}
As shown in previous sections, the amplitude encoding is advantageous when the quantum circuit width (i.e. the number of qubits) is considered while the qubit encoding is advantageous when the quantum circuit depth is considered. These two encoding schemes represent the extreme ends of the quantum circuit complexities for loading classical data into a quantum system. In this section, we introduce simple hybrid encoding methods to compromise the quantum circuit complexity between these two extreme ends.
In essence, the hybrid encoding implements the amplitude encoding to a number of independent blocks of qubits in parallel. Let us denote the number of qubits in each independent block that amplitude-encodes classical data by $m$. Then each block can encode $O(2^m)$ classical data. Let us also denote that there are $b$ such blocks of $m$ qubits by $b$. Then the quantum system of $b$ blocks contain $b2^m$ classical data. 
The first hybrid encoding, which we refer to as hybrid direct encoding (HDE), can be expressed as
\begin{equation}
\label{eq:hybrid1}
    U_{\phi}(x) : x \in \mathbb{R}^{N} \rightarrow \ket{\phi(x)} = \bigotimes_{j=1}^{b} \left( \frac{1}{\|x\|_j} \sum_{i=1}^{2^m} x_{ij} \ket{i}_j \right).
\end{equation}
Note that each block can have a different normalization constant, and hence the amplitudes may not be a faithful representation of the data unless the normalization constant have similar values. To circumvent this problem, we also introduce hybrid angle encoding (HAE), which can be expressed as
\begin{align}
\label{eq:hybrid2}
    &\ket{\phi(x)} \nonumber \\
    & = \bigotimes_{k=1}^{b} \left(\sum_{i=1}^{2^m} \prod_{j=0}^{m-1} \cos^{1-\mathtt{i}_j}\left(x_{g(j),k}\right)\sin^{\mathtt{i}_j}\left(x_{g(j),k}\right) \ket{i}_k \right),
\end{align}
where $\mathtt{i}\in\lbrace 0,1\rbrace^m$ is the binary representation of $i$ with $\mathtt{i}_{j}$ being the $j+1$th bit of the bit string, $x_{j,k}$ represents the $j$th element of the data assigned to the $k$th block of qubits, and $g(j)=2^j+\sum_{l=0}^{j-1}\mathtt{i}_l2^l$. In this case, having $b$ block of $m$ qubits allows $b(2^m-1)$ classical data to be encoded. The performance of these hybrid encoding methods will be compared in Sec.~\ref{sec:simulation}.

Since the hybrid methods are parallelized, the quantum circuit depth is reduced to $O(2^m)$ where $m<N$, while the number of qubits is $O(mN/2^m)$. Therefore the hybrid encoding algorithms use fewer number of qubits than the qubit encoding and use shallower quantum circuit depth than the amplitude encoding. Finding the best trade-off between the quantum circuit width and depth (i.e. the choice of $m$) depends on the specific details of given quantum hardware.


\section{Benchmark variables}
\label{sec:bmv}

\subsection{Ansatz}
An important step in the construction of a QCNN model is the choice of ansatz. In general, the QCNN structure is flexible to use an arbitrary two-qubit unitary operation at each convolutional filter and each pooling step. However, we constrain our design such that all convolutional filters use the same ansatz, and the same applies to all pooling operations (but differ from convolutional filters). We later show that the QCNN with fixed ansatz provides excellent results for the benchmarking datasets. While using different ansatz for all filters can be an interesting attempt for further improvements, this will increase the number of parameters to be optimized. 

In the following, we introduce a set of convolutional and pooling ansatz (i.e. parameterized quantum circuit templates) used in our QCNN models.

\subsubsection{Convolution filter}
\begin{figure*}[t]
    \centering
    \includegraphics[width=0.85\textwidth]{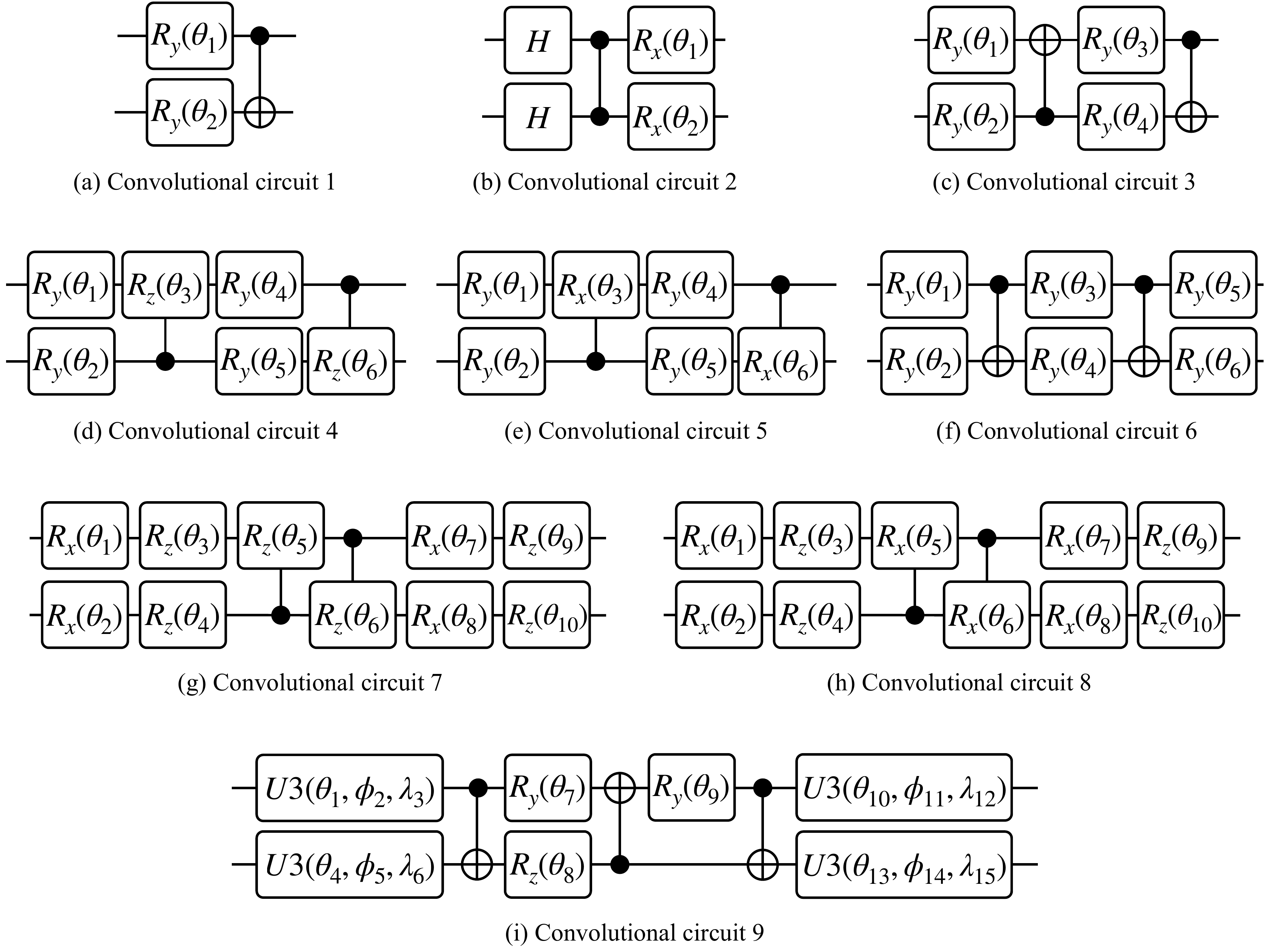}
    \caption{Parameterized quantum circuits used in the convolutional layer. $R_i(\theta)$ is a rotation around the $i$ axis of the Bloch sphere by an angle of $\theta$, and $H$ is the Hadamard gate. $U3(\theta,\phi,\lambda)$ is an arbitrary single-qubit gate that can be expressed as $U3(\theta,\phi,\lambda) = R_z(\phi)R_x(-\pi/2)R_z(\theta)R_x(\pi/2)R_z(\lambda)$.}
    \label{fig:convolution}
\end{figure*}

parameterized quantum circuits for convolutional layers in QCNN are composed of different configurations of single-qubit and two-qubit gate operations. Most circuit diagrams in Fig.~\ref{fig:convolution} are inspired by past studies. For instance, circuit 1 is used as the parameterized quantum circuit for training a tree tensor network (TTN)~\cite{grant_hierarchical_2018}. Circuits 2, 3, 4, 5, 7, and 8 are taken from the work by Sim et al. ~\cite{Sim_expressibility} which includes the analysis on expressibility and entangling capability of four-qubit parameterized quantum circuits. We modified these quantum circuits to two-qubit forms to utilize them as building blocks of the convolutional layer, which always consists of two qubits. Circuits 7 and 8 are reduced versions of circuits that recorded the best expressibility in the study. Circuit 2 is a two-qubit version of the quantum circuit that exhibited the best entangling capability. Circuits 3, 4 and 5 are drawn from circuits that have balanced significance in both expressibility and entangling capability. Circuit 6 is developed as a proper candidate of two-body Variational Quantum Eigensolver (VQE) entangler in Ref.~\cite{PhysRevLett.122.230401}. This circuit is also known to be able to implement an arbitrary $SO(4)$ gate~\cite{DBLP:journals/qic/WeiD12}. In fact, a total VQE entangler can be constructed by linearly arranging the $SO(4)$ gates throughout input qubits. Since this structure is similar to the structure of convolutional layers in QCNN, the $SO(4)$ gate would be a great candidate to be used in the convolution layer. Circuit 9 represents the parameterization of an arbitrary $SU(4)$ gate~\cite{PhysRevA.69.032315,maccormack_branching_2020}.

\subsubsection{Pooling}

\begin{figure}[t]
    \centering
    \includegraphics[width=0.44\columnwidth]{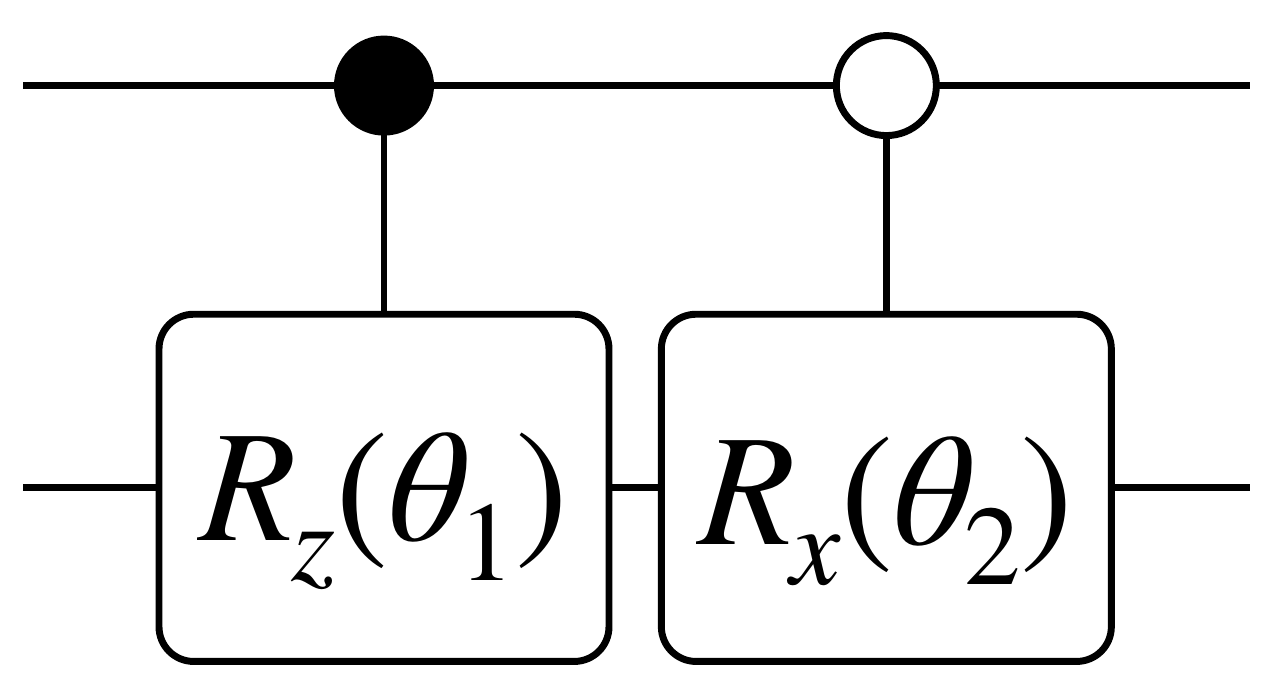}
    \caption{Parameterized quantum circuit used in the pooling layer. The pooling layer applies two controlled rotations $R_z(\theta_1)$ and $R_x(\theta_2)$, respectively, each activated when the control qubit is 1 (filled circle) or 0 (open circle). The control (first) qubit is traced out after the gate operations in order to reduce the dimension.}
    \label{fig:pooling}
\end{figure}

The pooling layer applies parameterized quantum gates to two qubits and traces out one of the qubits to reduce the two-qubit states to one-qubit states. Similar to the choice of ansatz for the convolutional filter, there exists a variety of choices of two-qubit circuits of the pooling layer. In this work, we choose a simple form of a two-qubit circuit consisting of two free parameters for the pooling layer. The circuit is shown in Fig.~\ref{fig:pooling}.

Application of the parameterized gates in the pooling step in conjunction with the convolutional circuit 9 might be redundant since it is already an arbitrary $SU(4)$ gate. Thus for the convolutional circuit 9, we test two QCNN constructions, with and without the parameterized two-qubit circuit in the pooling layer. In the latter, the pooling layer only consists of tracing out one qubit.

\subsection{Cost function}

The variational parameters of the ansatz are updated to minimize the cost function calculated on the training data set. In this benchmark study, we test the performance of QCNN models with two different cost functions, namely the mean squared error and the cross-entropy loss.

\subsubsection{Mean Squared Error}
Before training QCNN, we map original class labels of $\lbrace 0,1\rbrace$ to $\lbrace 1,-1\rbrace$ respectively to associate them with the eigenvalues of the qubit observables. Then the mean squared error (MSE) between predictions and class labels becomes
\begin{equation}
    C(\boldsymbol{\theta}) = \frac{1}{N} \sum_{i=1}^{N} (\hat{M_{z}}(\psi_i(\boldsymbol{\theta})) - \tilde{y}_i)^2,
\end{equation} 
where $\hat{M_{z}}(\psi_i)=\langle\psi_i|\sigma_z|\psi_i\rangle$ is the Pauli-Z expectation value of one qubit state extracted from QCNN for $i$th training data, and $\tilde{y}_i$ $\in \{1, -1\}$ is the label of the corresponding training data (i.e. $\tilde{y}_i = 1-2y_i$). Since QCNN performs a single qubit measurement in the $Z$ basis, the final state can be thought of as a mixed state $a_i\ketbra{0}{0}+b_i\ketbra{1}{1}$. Then minimizing the cost function above with respect to $\boldsymbol{\theta}$ would correspond to forcing $a_i$ to be as larger as possible than $b_i$ if the $i$th training data is labelled $0$, and vice versa if it is labelled $1$.

\subsubsection{Cross-Entropy Loss}
Cross-entropy loss is widely used in training classical neural networks. It measures the performance of a classification model whose output is a probability between 0 and 1. Due to the probabilistic property of quantum mechanics, one could consider the cross-entropy loss by considering probabilities of measuring computational basis states in the single-qubit measurement of QCNN. The cross-entropy loss for the $i$th training data can be expressed as
\begin{align}
    C(\boldsymbol{\theta}) = \sum_{i=1}^{N}\Big{[}&y_i\log\left(\Pr[\psi_i(\boldsymbol{\theta}) = 1]\right) \nonumber \\ 
    & + (1-y_i)\log\left(\Pr[\psi_i(\boldsymbol{\theta}) = 0]\right)\Big{]},
\end{align}
where $y_i \in \{0, 1\}$ is the class label and $\Pr[\psi_i(\boldsymbol{\theta}) = y_i]$ is the probability of measuring the computational basis state $\ket{y_i}$ from the QCNN circuit.

\subsection{Classical data pre-processing}

The size of quantum circuits that can be reliably executed on NISQ devices is limited due to the noise and technical challenges of building quantum hardware. Thus the encoding schemes for high dimensional data usually require the number of qubits that are beyond the current capabilities of quantum devices. Therefore, classical dimensionality reduction techniques will be useful in the near-term application of quantum machine learning techniques. In this work, we pre-process data with three classical dimensionality reduction techniques, namely bilinear interpolation, principal component analysis (PCA)~\cite{PCA} and autoencoding (AutoEnc)~\cite{Goodfellow-et-al-2016}. For the simulation presented in the following section, amplitude encoding is used only with bilinear interpolation while all other encoding schemes are tested with PCA and autoencoding. Bilinear interpolation and PCA are carried out by utilizing \texttt{tf.image.resize} from TensorFlow and \texttt{sklearn.decomposition.PCA} from scikit-learn, respectively. Autoencoders are capable of modelling complex non-linear functions, while PCA is a simple linear transformation with cheaper and faster computation. Since the pre-processing step should not produce too much computational resource overhead or result in overfitting, we train a simple autoencoder with one hidden layer. The data in the latent space (i.e. hidden layer) is then fed to quantum circuits.

\section{Simulation}
\label{sec:simulation}

\subsection{QCNN results overview}

This section reports the classical simulation results of the QCNN algorithm for binary classification carried out with Pennylane~\cite{bergholm2020pennylane}. The test is performed with two standard datasets, namely MNIST and Fashion MNIST, under various conditions as described in the previous section. Note that the MNIST and Fashion MNIST datasets are 28$\times$28 image data, each with ten classes. Our benchmark focuses on binary classification, and hence we select classes 0 and 1 for both datasets.

The variational parameters in the QCNN ansatze are optimized by minimizing the cost function with an optimizer provided in Pennylane~\cite{bergholm2020pennylane}. In particular, we tested Adam~\cite{kingma2017adam} and Nesterov moment~\cite{Nesterov1983AMF} optimization algorithms. At each iteration $t$, we create a small batch by randomly selecting data from the training set. 
Compared to training on the full data set, training on the mini-batch not only reduces simulation time but also helps the gradients to escape from local minima. For both Adam and Nesterov moment optimizers, the batch size was 25 and the learning rate was 0.01. We also fixed the number of iterations to be 200 to speed up the training process. Note that the training can alternatively be stopped at different conditions, such as when the validation set accuracy does not increase for a predetermined number of consecutive runs~\cite{grant_hierarchical_2018}. The number of training (test) data are 12665 (2115) and 12000 (2000) for MNIST and fashion MNIST datasets, respectively. 

Table~\ref{tab:MNIST_ce} and~\ref{tab:FMNIST_ce} show the mean classification accuracy and one standard deviation obtained from five instances with random initialization of parameters. The number of random initialization is chosen to be the same as that of Ref.~\cite{grant_hierarchical_2018}. The results are obtained for various QCNN models of different convolutional and pooling circuits and data encoding strategies. When benchmarking with the hybrid encoding schemes (i.e. HDE and HAE), we used two blocks of four qubits, which results in having 32 and 30 features encoded in 8 qubits, respectively.  For all results presented here, training is done with the cross-entropy loss. Similar results are obtained when MSE is used for training, and we present the MSE results in Appendix~\ref{sec:AppC}. Here we only report the classification results obtained with the Nesterov optimizer, since it consistently provided better convergence. The ansatze in the table are listed in the same order as the list of convolutional circuits shown in Fig.~\ref{fig:convolution}. The last row of the table (i.e. Ansatz 9b) contains the results when the QCNN circuit only consists of the convolutional circuit 9 without any unitary gates in the pooling step.

\begin{table*}[ht]
\centering
\begin{tabular}{@{}ccccccc@{}}
\toprule
\multicolumn{2}{c}{}        & \multicolumn{5}{c}{Classification Accuracy} \\ \midrule
\multirow{2}{*}{Ansatz} &
  \multirow{2}{*}{\begin{tabular}[c]{@{}c@{}}\# of \\ params\end{tabular}} &
  \multirow{2}{*}{Amplitude} &
  \multirow{2}{*}{Qubit} &
  \multirow{2}{*}{Dense} &
  \multirow{2}{*}{HDE} & 
  \multirow{2}{*}{HAE} \\
 &                        &           &           &          &  &         \\ \midrule
1 & \multicolumn{1}{c|}{12} &  $96.8\pm 5.3$ &  \begin{tabular}[c]{@{}c@{}}$98.0\pm 0.4$ \\ $91.4\pm 2.3$\end{tabular}         &   \begin{tabular}[c]{@{}c@{}} $97.6\pm 1.1$    \\ $88.4\pm 9.2$\end{tabular}        &   \begin{tabular}[c]{@{}c@{}}  $68.7 \pm 5.1$   \\ $88.4 \pm 2.6$ \end{tabular}   &   \begin{tabular}[c]{@{}c@{}} $97.9 \pm 0.3$    \\ $77.7 \pm 6.0$ \end{tabular}   \\
2 & \multicolumn{1}{c|}{12} & $94.5\pm 3.1$ & \begin{tabular}[c]{@{}c@{}} $98.2\pm 4.5$ \\ $98.2\pm 6.6$\end{tabular} & \begin{tabular}[c]{@{}c@{}} $98.2 \pm 0.5$    \\ $85.6 \pm 4.5$\end{tabular}     &
\begin{tabular}[c]{@{}c@{}} $62.2 \pm 3.2$    \\ $93.4 \pm 5.4$ \end{tabular} &
\begin{tabular}[c]{@{}c@{}} $94.7 \pm 2.1$    \\ $80.0 \pm 4.0$ \end{tabular} \\
3 & \multicolumn{1}{c|}{18} & $93.8\pm 4.4$ & \begin{tabular}[c]{@{}c@{}} $\bold{98.5}\pm 0.2$ \\ $93.3\pm 3.8$\end{tabular} &   \begin{tabular}[c]{@{}c@{}} $96.9\pm 1.6$ \\ $95.8\pm 1.7$\end{tabular}        &
\begin{tabular}[c]{@{}c@{}} $76.4 \pm 2.7$    \\ $95.3 \pm 3.4$ \end{tabular}&
\begin{tabular}[c]{@{}c@{}} $98.1 \pm 0.2$    \\ $84.4 \pm 0.7$ \end{tabular}\\
4 & \multicolumn{1}{c|}{24} & $97.8\pm 2.4$ & \begin{tabular}[c]{@{}c@{}} $98.2\pm 0.4  $ \\ $\bold{98.5}\pm 1.2$\end{tabular} &  \begin{tabular}[c]{@{}c@{}} $98.2\pm 0.4$ \\ $\bold{97.2} \pm 1.1$\end{tabular}         &
\begin{tabular}[c]{@{}c@{}} $70.2 \pm 1.3$    \\ $96.6 \pm 1.0$ \end{tabular}&
\begin{tabular}[c]{@{}c@{}} $98.0 \pm 0.3$    \\ $\bold{90.4} \pm 3.8$ \end{tabular}\\
5 & \multicolumn{1}{c|}{24} & $96.7\pm 2.1$ & \begin{tabular}[c]{@{}c@{}} $98.3\pm 0.4$ \\ $94.9\pm 2.1$\end{tabular} &  \begin{tabular}[c]{@{}c@{}} $98.1\pm 0.5$ \\ $96.0\pm 1.3$\end{tabular}         &
\begin{tabular}[c]{@{}c@{}} $72.6 \pm 5.7$    \\ $93.5 \pm 1.3$ \end{tabular}&
\begin{tabular}[c]{@{}c@{}} $98.0 \pm 0.1$    \\ $86.5 \pm 5.0$ \end{tabular}\\
6 & \multicolumn{1}{c|}{24} & $97.2\pm 2.2$ & \begin{tabular}[c]{@{}c@{}} $98.1\pm 0.4$ \\ $97.7\pm 1.0$\end{tabular} &   \begin{tabular}[c]{@{}c@{}} $98.1\pm 0.3$ \\ $93.4\pm 0.5$\end{tabular}          &   \begin{tabular}[c]{@{}c@{}} $77.4\pm 1.7$ \\ $97.0\pm 2.0$\end{tabular}    &
\begin{tabular}[c]{@{}c@{}} $\bold{98.3}\pm 0.2$ \\ $86.9\pm 7.3$\end{tabular} \\  
7 & \multicolumn{1}{c|}{36} & $98.3\pm 2.2$ & \begin{tabular}[c]{@{}c@{}} $98.2\pm 0.3$ \\ $93.7\pm 4.5$\end{tabular} &   \begin{tabular}[c]{@{}c@{}} $\bold{98.7}\pm 2.4$ \\ $95.1\pm 1.6$\end{tabular}      & \begin{tabular}[c]{@{}c@{}} $74.6 \pm 3.2$    \\ $97.2 \pm 2.2$ \end{tabular}     &
\begin{tabular}[c]{@{}c@{}} $98.2 \pm 0.1$    \\ $90.2 \pm 3.0$ \end{tabular}\\
8 & \multicolumn{1}{c|}{36} & $98.1\pm 0.7$ & \begin{tabular}[c]{@{}c@{}} $98.3\pm 0.4$ \\ ${96.9}\pm 2.4$\end{tabular} &   \begin{tabular}[c]{@{}c@{}} $\bold{98.7}\pm 0.1$ \\ $95.4\pm 2.8$\end{tabular}         &\begin{tabular}[c]{@{}c@{}} $\bold{79.7} \pm 1.6$    \\ $96.6 \pm 1.7$ \end{tabular}      &
\begin{tabular}[c]{@{}c@{}} $\bold{98.3} \pm 0.1$    \\ $89.1 \pm 2.6$ \end{tabular}\\
9a & \multicolumn{1}{c|}{51} & $\bold{98.4}\pm 0.2$ & \begin{tabular}[c]{@{}c@{}} $98.4\pm 0.5$ \\  $96.4\pm 2.3$\end{tabular} &    \begin{tabular}[c]{@{}c@{}} $\bold{98.7} \pm 0.4$ \\  ${96.7} \pm 1.4$\end{tabular}        &
\begin{tabular}[c]{@{}c@{}} $78.1 \pm 2.8$    \\ $97.8 \pm 2.2$ \end{tabular}&
\begin{tabular}[c]{@{}c@{}} ${98.2} \pm 0.2$    \\ $87.0 \pm 5.3$ \end{tabular}\\  
9b & \multicolumn{1}{c|}{45} & $98.3\pm 0.2$ & \begin{tabular}[c]{@{}c@{}} ${97.7}\pm 0.6$ \\ ${96.6}\pm 2.2$\end{tabular} &   \begin{tabular}[c]{@{}c@{}} $98.3\pm 0.5$ \\ $96.5\pm 1.7$\end{tabular}       & 
\begin{tabular}[c]{@{}c@{}} $77.4\pm 2.5$ \\ $\textbf{98.0}\pm 1.2$\end{tabular}&
\begin{tabular}[c]{@{}c@{}} $98.1\pm 0.1$ \\ $88.5\pm 2.8$\end{tabular}\\\bottomrule
\end{tabular}
\caption{\label{tab:MNIST_ce}Mean accuracy and one standard deviation of the classification for 0 and 1 in the MNIST dataset when the model is trained with the cross-entropy loss. The mean and the standard deviation are obtained from five repetitions with random initialization of parameters. The first column shows the ansatz label. The second column shows the total number of parameters that are subject to optimization. For qubit, dense and hybrid encoding, two rows indicate the values obtained with different classical data pre-processing, namely PCA and autoencoding, respectively. The best result under each quantum data encoding method is written in bold.}
\end{table*}

\begin{table*}[ht]
\centering
\begin{tabular}{@{}ccccccc@{}}
\toprule
\multicolumn{2}{c}{}        & \multicolumn{5}{c}{Classification Accuracy} \\ \midrule
\multirow{2}{*}{Ansatz} &
  \multirow{2}{*}{\begin{tabular}[c]{@{}c@{}}\# of \\ params\end{tabular}} &
  \multirow{2}{*}{Amplitude} &
  \multirow{2}{*}{Qubit} &
  \multirow{2}{*}{Dense} &
  \multirow{2}{*}{HDE} & 
  \multirow{2}{*}{HAE} \\
 &                        &           &           &          &  &         \\ \midrule
1 & \multicolumn{1}{c|}{12} &  $90.9\pm 2.0$ &  \begin{tabular}[c]{@{}c@{}}$83.1\pm 3.3$ \\ $87.3\pm 4.2$\end{tabular}         &   \begin{tabular}[c]{@{}c@{}} $85.0\pm 3.6$    \\ $84.7\pm 7.2$\end{tabular}        &   \begin{tabular}[c]{@{}c@{}}  $64.4 \pm 3.2$   \\ $90.5 \pm 1.3$ \end{tabular}   &   \begin{tabular}[c]{@{}c@{}} $83.9 \pm 1.7$   \\ $82.7 \pm 4.6$ \end{tabular}   \\
2 & \multicolumn{1}{c|}{12} & $88.2\pm 3.8$ & \begin{tabular}[c]{@{}c@{}} $87.2\pm 4.6$ \\ $86.6\pm 4.6$\end{tabular} & \begin{tabular}[c]{@{}c@{}} $82.2 \pm 2.1$    \\ $86.9 \pm 7.7$\end{tabular}     & \begin{tabular}[c]{@{}c@{}} $63.1 \pm 1.4$    \\ $86.0 \pm 4.7$\end{tabular} &     \begin{tabular}[c]{@{}c@{}} $84.3 \pm 4.1$    \\ $82.1 \pm 5.0$\end{tabular}    \\
3 & \multicolumn{1}{c|}{18} & $90.1\pm 2.7$ & \begin{tabular}[c]{@{}c@{}} ${87.7}\pm 3.6$ \\ $88.4\pm 7.4$\end{tabular} &   \begin{tabular}[c]{@{}c@{}} $87.2\pm 3.0$ \\ $88.8\pm 2.2$\end{tabular}        &  \begin{tabular}[c]{@{}c@{}} $65.5\pm 1.3$ \\ $91.7\pm 1.5$\end{tabular}    &    
\begin{tabular}[c]{@{}c@{}} $85.5\pm 1.2$ \\ $88.1\pm 4.0$\end{tabular}\\
4 & \multicolumn{1}{c|}{24} & $89.1\pm 2.2$ & \begin{tabular}[c]{@{}c@{}} $87.2\pm 2.4  $ \\ $91.6\pm 3.4$\end{tabular} &  \begin{tabular}[c]{@{}c@{}} $89.7\pm 1.6$ \\ ${91.8} \pm 1.7$\end{tabular}         &  \begin{tabular}[c]{@{}c@{}} $64.7\pm 1.9$ \\ ${91.2} \pm 0.5$\end{tabular}&
\begin{tabular}[c]{@{}c@{}} $84.7\pm 1.1$ \\ ${88.4} \pm 4.4$\end{tabular}\\
5 & \multicolumn{1}{c|}{24} & $90.7\pm 1.1$ & \begin{tabular}[c]{@{}c@{}} $86.3\pm 2.9$ \\ $91.9\pm 1.4$\end{tabular} &  \begin{tabular}[c]{@{}c@{}} $87.6\pm 2.3$ \\ $\textbf{93.7}\pm 1.4$\end{tabular}         &     \begin{tabular}[c]{@{}c@{}} $64.1\pm 2.4$ \\ $92.4\pm 1.3$\end{tabular}&
\begin{tabular}[c]{@{}c@{}} $84.5\pm 1.7$ \\ $85.9\pm 3.9$\end{tabular}\\
6 & \multicolumn{1}{c|}{24} & $90.4\pm 1.7$ & \begin{tabular}[c]{@{}c@{}} $87.9\pm 3.9$ \\ $93.6\pm 1.6$\end{tabular} &   \begin{tabular}[c]{@{}c@{}} $88.7\pm 2.4$ \\ $90.2\pm 1.9$\end{tabular}          &   \begin{tabular}[c]{@{}c@{}} $65.7\pm 0.9$ \\ $94.3\pm 1.2$\end{tabular}    &
\begin{tabular}[c]{@{}c@{}} $87.7\pm 2.6$ \\ $88.5\pm 2.6$\end{tabular} \\  
7 & \multicolumn{1}{c|}{36} & $88.2\pm 1.4$ & \begin{tabular}[c]{@{}c@{}} $87.1\pm 3.8$ \\ $92.2\pm 0.5$\end{tabular} &   \begin{tabular}[c]{@{}c@{}} ${88.7}\pm 2.9$ \\ $92.7\pm 2.2$\end{tabular}      &     \begin{tabular}[c]{@{}c@{}} $66.3\pm 0.9$ \\ $93.8\pm 0.8$\end{tabular} & 
\begin{tabular}[c]{@{}c@{}} $86.7\pm 1.8$ \\ $89.7\pm 2.9$\end{tabular}\\
8 & \multicolumn{1}{c|}{36} & $89.9\pm 1.9$ & \begin{tabular}[c]{@{}c@{}} $\bold{89.8}\pm 0.9$ \\ ${91.6}\pm 3.0$\end{tabular} &   \begin{tabular}[c]{@{}c@{}} $88.7\pm 2.6$ \\ $92.2\pm 2.6$\end{tabular}         &\begin{tabular}[c]{@{}c@{}} $\bold{66.5}\pm 0.3$ \\ $93.2\pm 1.5$\end{tabular}      & 
\begin{tabular}[c]{@{}c@{}} $86.2\pm 1.5$ \\ $\textbf{91.6}\pm 3.6$\end{tabular}\\
9a & \multicolumn{1}{c|}{51} & $\textbf{91.3}\pm 2.3$ & \begin{tabular}[c]{@{}c@{}} $88.9\pm 2.7$ \\  $92.4\pm 2.6$\end{tabular} &    \begin{tabular}[c]{@{}c@{}} ${89.2} \pm 2.0$ \\  ${92.1} \pm 1.3$\end{tabular}        &   \begin{tabular}[c]{@{}c@{}} $65.0\pm 1.3$ \\ $\textbf{94.3}\pm 1.6$\end{tabular}   &
\begin{tabular}[c]{@{}c@{}} $88.6\pm 2.0$ \\ $88.3\pm 3.7$\end{tabular}\\  
9b & \multicolumn{1}{c|}{45} & ${88.8}\pm 1.6$ & \begin{tabular}[c]{@{}c@{}} ${88.0}\pm 2.5$ \\ $\textbf{94.1}\pm 1.1$\end{tabular} &   \begin{tabular}[c]{@{}c@{}} $\bold{90.2}\pm 0.5$ \\ $92.7\pm 1.5$\end{tabular}       & 
\begin{tabular}[c]{@{}c@{}} $64.7\pm 2.8$ \\ $92.9\pm 1.3$\end{tabular}&
\begin{tabular}[c]{@{}c@{}} $\bold{88.9}\pm 1.6$ \\ $87.1\pm 3.6$\end{tabular}\\\bottomrule
\end{tabular}
\caption{\label{tab:FMNIST_ce}Mean accuracy and one standard deviation of the classification for 0 (t-shirt/top) and 1 (trouser) in the Fashion MNIST dataset when the model is trained with the cross-entropy loss. The mean and the standard deviation are obtained from five repetitions with random initialization of parameters. The first column shows the ansatz label. The second column shows the total number of parameters that are subject to optimization. For qubit, dense and hybrid encoding, two rows indicate the values obtained with different classical data pre-processing, namely PCA and autoencoding, respectively. The best result under each quantum data encoding method is written in bold.}
\end{table*}

The simulation results show that all ansatze perform reasonably well, while the ones with more number of free parameters tend to produce higher score. Since all ansatze perform reasonably well, one may choose to use the ansatz with smaller number of free parameters to save the training time. For example, by choosing ansatz 4 and amplitude encoding, one can achieve $97.8\%$ classification accuracy while using only 24 free parameters total instead of achieving $98.4\%$ at the cost of increasing the number of free parameters to 51. It is also important to note that most of the results obtained with the hybrid data encoding and PCA is considerably worse than the others. This is due to the normalization problem discussed in Sec.~\ref{sec:hybrid_enc}, which motivated the development of the hybrid angle encoding. We observed that the normalization problem is negligible in the case with autoencoding. The simulation results clearly demonstrates that the normalization problem is resolved by the hybrid angle encoding as expected, and reasonably good results can be obtained with this method. For MNIST, HAE with PCA provides the best solution among the hybrid encoding schemes on average. On the other hand, for Fashion MNIST, HDE with autoencoding provides the best solution among the hybrid encoding schemes on average.

In the following, we also compare the two classical dimensionality reduction methods by presenting the overall mean accuracy and standard deviation values obtained by averaging over all ansatze and the random initializations. The average values are presented in Tab.~\ref{tab:PCAvsAE}. As discussed before, the HDE with PCA does not perform well for both datasets due to the data normalization issue. Besides this case, interestingly, PCA works better than autoencoding for MNIST data, and vice versa for Fashion MNIST data, thereby suggesting that the choice of the classical pre-processing method should be data-dependent.

\begin{table}[ht]
\centering
\begin{tabular}{@{}cccccc@{}}
\toprule
 &  & Qubit & Dense & HDE & HAE \\ \midrule
\multicolumn{1}{c|}{\multirow{2}{*}{MNIST}} & \multicolumn{1}{c|}{PCA} & $98.0\pm 1.5$ & $98.0\pm 1.0$ & $73.7\pm 3.0$ & $97.8\pm 0.4$ \\ \cmidrule(lr){2-2}
\multicolumn{1}{c|}{} & \multicolumn{1}{c|}{AutoEnc} & $95.4\pm 3.6$ & $94.2\pm 4.6$ & $95.4\pm 2.3$ & $86.1\pm 4.1$ \\ \midrule
\multicolumn{1}{c|}{\multirow{2}{*}{\begin{tabular}[c]{@{}c@{}}Fashion\\ MNIST\end{tabular}}} & \multicolumn{1}{c|}{PCA} & $87.3\pm 3.7$ & $87.7\pm 3.2$ & $65.0\pm 1.6$ & $86.1\pm 1.9$ \\ \cmidrule(lr){2-2}
\multicolumn{1}{c|}{} & \multicolumn{1}{c|}{AutoEnc} & $91.0\pm 4.1$ & $90.6\pm 4.4$ & $92.0\pm 1.5$ & $87.2\pm 3.8$ \\ \bottomrule
\end{tabular}
\caption{\label{tab:PCAvsAE}Comparison of the classical dimensionality reduction methods for angle encoding, dense encoding, and hybrid encoding. For each encoding scheme, classification results from all instances (i.e. various ansatze and random initialization of parameters) are averaged out to produce the mean and standard deviation.}
\end{table}

\begin{figure*}[t]
    \centering
    \includegraphics[width=0.9\textwidth]{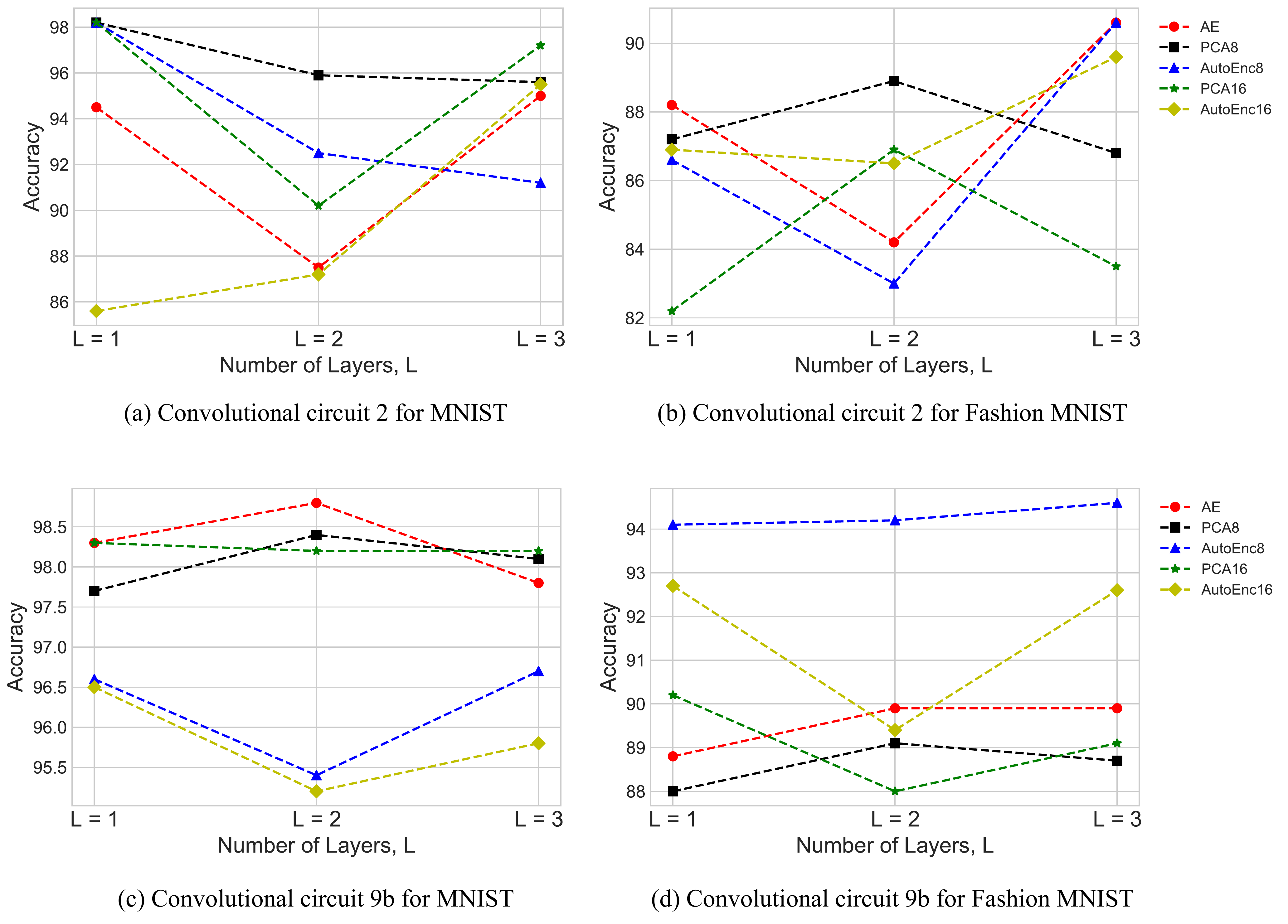}
    \caption{\label{fig:layer}Classification accuracy vs. the number of convolutional filters $L$ for MNIST and Fashion MNIST datasets. The number of filters in each layer is set to be equal, i.e. $l_1=l_2=l_3=L$ (see Fig.~\ref{fig:1} for the definition of $l_i$). The simulation is carried out with two ansatze, circuit 2 and circuit 9b, and five encoding schemes, amplitude encoding (AE), qubit encoding with PCA (PCA8) and with autoencoding (AutoEnc8), and dense encoding with PCA (PCA16) and with autoencoding (AutoEnc16).}
\end{figure*}

Finally, we also examine how classification performance improves as the number of convolutional filters in each layer increases. For simplicity, we set the number of convolutional filters in each layer to be same, i.e $l_1=l_2=l_3=L$ (see Fig.~\ref{fig:1} for the definition of $l_i$). Without loss of generality, we pick two ansatze and five encodings. For ansatze, we choose the one with the smallest number of free parameters and another with arbitrary $SU(4)$ operations. These are circuit 2 and circuit 9b, and they use 12 and 45 parameters total, respectively. For data encoding, we tested amplitude, qubit, and dense encoding. The qubit and dense encoding are further grouped under two different classical dimensionality reduction techniques, PCA and autoencoding. Since the qubit and dense encoding load 8 and 16 features, respectively, we label them as PCA8, AutoEnc8, PCA16, and AutoEnc16 based on the number of features and the  dimensionality reduction techniques. The classification accuracies for $L=\lbrace 1,2,3\rbrace$ are plotted in Fig.~\ref{fig:layer}. The simulation results show that in some cases, the classification accuracy can be improved by increasing the number of convolutional filters. For example, the classification accuracy for MNIST data can be improved from about $86\%$ to $96\%$ when circuit 2 and dense encoding with autoencoding are used. For Fashion MNIST, the classification accuracy is improved from about $88\%$ to $90\%$ when circuit 2 and amplitude encoding are used, and from about $86\%$ to $90\%$ when circuit 2 and qubit encoding with PCA is used. However, we do not observe general trend with respect to the number of convolutional filters. In particular, the relationship between the classification accuracy and $L$ is less obvious for circuit 9b. We speculate that this attributes to the fact that circuit 9b implements an arbitrary $SU(4)$, which is an arbitrary two-qubit gate, and hence repetitive application of an arbitrary $SU(4)$ is redundant.

\subsection{Boundary conditions of the QCNN circuit}

\begin{figure}[ht]
    \centering
    \includegraphics[width=0.7\columnwidth]{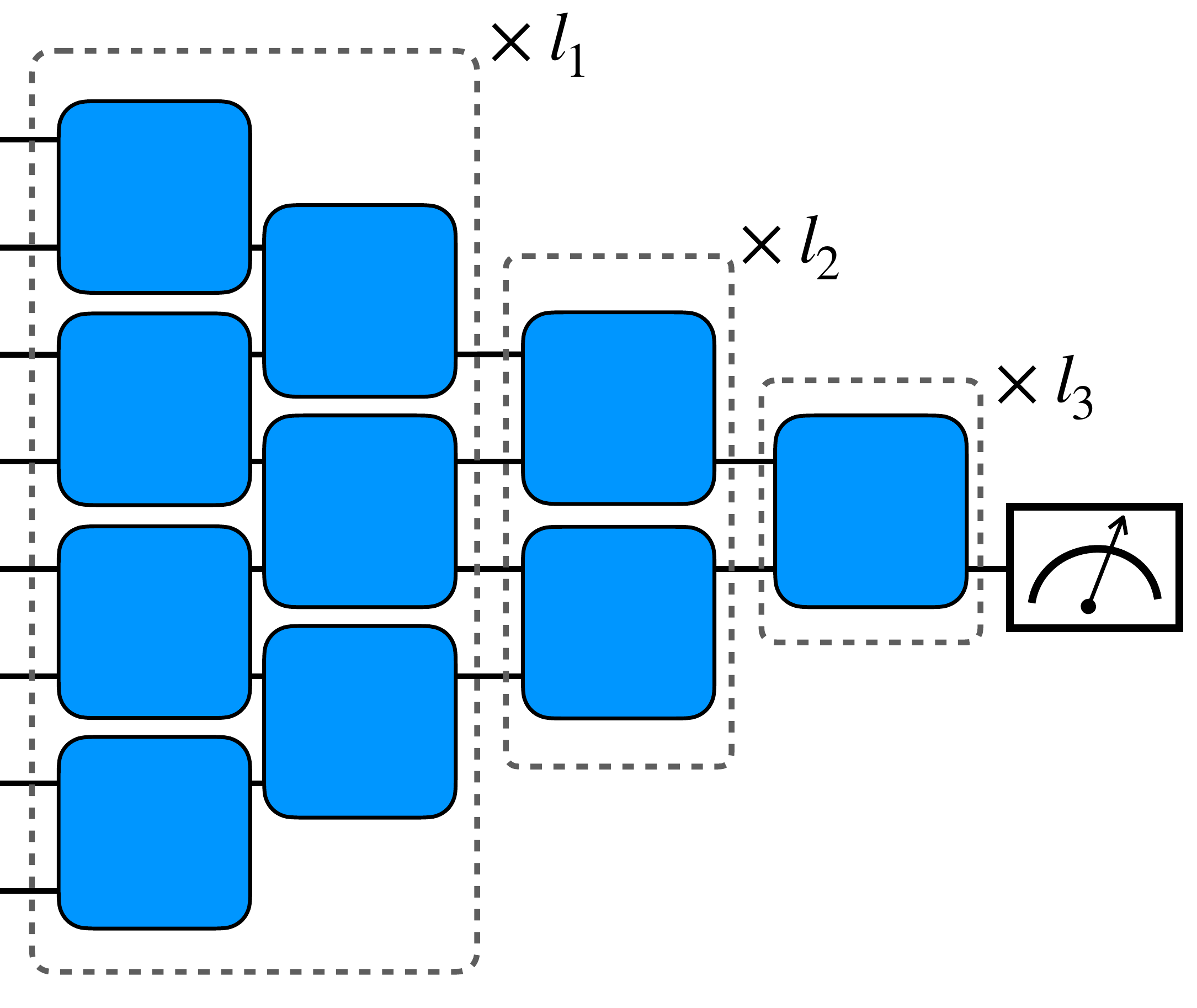}
    \caption{\label{fig:hwefficient}A QCNN circuit with the open-boundary condition and no gate operations for pooling. In this case, the QCNN circuit can be constructed with nearest-neighbour qubit interactions only.}
\end{figure}

The general structure of QCNN shown in Fig.~\ref{fig:1} uses two-qubit gates between the first (top) and last (bottom) qubits, which can be thought of as imposing periodic boundary condition. One may notice that all-to-all connectivity can be established even without connecting the boundaries. Thus we tested the classification performance of a QCNN architecture without having the two-qubit gates to close the loop. We refer to this case as the open boundary QCNN. Without loss of generality, we tested QCNNs with two different ansatz, the convolutional circuit 2 (Ansatz 2 in Tabs.~\ref{tab:MNIST_ce} and \ref{tab:FMNIST_ce}) which uses the smallest number of free parameters and the convolutional circuit 9 which implements an arbitrary $SU(4)$. In case of the latter, pooling was done without parameterized gates, and hence the ansatz is equivalent to ansatz 9b in Tabs.~\ref{tab:MNIST_ce} and \ref{tab:FMNIST_ce}. By imposing the open-boundary condition in conjunction with the ansatz 9b, one can modify the qubit arrangement of the QCNN circuit so as to use nearest-neighbour qubit interactions only. For an example of 8-qubit QCNN circuit, the modified structure is depicted in Fig.~\ref{fig:hwefficient}. Such design is particularly advantageous for NISQ devices that have limited physical qubit connectivity. For example, if one employs the qubit or the dense encoding method, the QCNN algorithm can be implemented with a 1-dimensional chain of physical qubits.

\begin{table*}[ht]
\centering
\begin{tabular}{@{}c|c|ccccc@{}} 
 \toprule
 & Ansatz &AE & PCA8 & PCA16 & AutoEnc8 & AutoEnc16  \\ 
 \midrule
\multirow{2}{*}{MNIST}& 2 & $90.4 \pm {2.5}$ & $98.0 \pm {0.3}$ & $96.3 \pm {3.2}$ & $97.7 \pm {0.1}$ & $86.4 \pm {4.6}$\\ 
& 9b & $98.0 \pm {0.4}$ & $98.1 \pm {0.2}$ & $97.8 \pm {0.3}$ & $96.8 \pm {0.7}$ &  $94.4 \pm {1.9}$\\
 \midrule
\multirow{2}{*}{\begin{tabular}[c]{@{}c@{}}Fashion\\ MNIST\end{tabular}} & 2 & $91.1 \pm 1.4$ & $86.2 \pm 0.5$ & $88.4 \pm 3.0$ & $83.3 \pm 4.6$ & $87.2 \pm 5.7$\\ 
& 9b & $90.1 \pm 1.4$ & $87.6 \pm 2.4$ & $89.1 \pm 1.9$ & $91.9 \pm 1.4$ &  $92.6 \pm 2.0$\\
 \bottomrule
\end{tabular}
\caption{\label{tab:open}Mean classification accuracy and one standard deviation of the classification for 0 and 1 in the benchmarking datasets when the QCNN circuit is constructed under the open boundary condition. Each column represents the results produced under a different encoding scheme with the numbers 8 and 16 indicates the qubit and dense encoding, respectively.}
\end{table*}

The simulation results are presented in Tab.~\ref{tab:open} for MNIST and Fashion MNIST datasets. These results are attained with one convolutional filters per layer, i.e. $l_1=l_2=l_3=1$. The simulation results demonstrate that for the case of two ansatze tested the classification performance between open- and periodic-boundary QCNN circuits are similar. Although the number of free parameters are the same under these conditions, depending on the specification of the quantum hardware such as the qubit connectivity, the open-boundary QCNN circuit can have shallower depth. The open-boundary circuit with ansatz 9b is even more attractive for NISQ devices since the convolutional operations can be done with only nearest-neighbour qubit interactions as mentioned above.

\subsection{Comparison to CNN}

We now compare the classification results of QCNN to that of classical CNN. Our goal is to compare the classification accuracy of the two given a similar number of parameters subject to optimization. To make a fair comparison between the two, we fix all hyperparameters of the two methods to be the same, except we used the Adam optimizer for CNN since it performed significantly better than the Nesterov moment optimizer. A detailed description of the classical CNN architecture is provided in Appendix~\ref{sec:AppB}.

It is important to note that CNN can be trained with such small number of parameters effectively only when the number of nodes in the input layer is small. Therefore, the CNN results are only comparable to that of the qubit and dense encoding cases which requires 8 and 16 classical input nodes, respectively. We designed four different CNN models with the number of free parameters being 26, 34, 44 and 56 to make them comparable to the QCNN models. In these cases, a dimensionality reduction technique must precede. For hybrid and amplitude encoding, which require relatively simpler data pre-processing, the number of nodes in the CNN input layer is too large to be trained with a small number of parameters as in QCNN.

Comparing the values in Tab.~\ref{tab:cnn} with the QCNN results, one can see that QCNN models perform better than their corresponding CNN models for the MNIST dataset. The same conclusion also holds for the Fashion dataset, except for the CNN models with 44 and 56 parameters that achieve similar performance as their corresponding QCNN models. Another noticeable result is that the QCNN models have considerably smaller standard deviations than the CNN models on average. This implies that the QCNN models not only achieve higher classification accuracy than the CNN models under similar training conditions but also are less sensitive to the random initialization of the free parameters.
\begin{table}[h]
\centering
\begin{tabular}{@{}ccccc@{}}
\toprule
 &  &  & \multicolumn{2}{c}{Classification Accuracy} \\ \midrule
\multirow{2}{*}{} & \multirow{2}{*}{\begin{tabular}[c]{@{}c@{}}\# of\\ params\end{tabular}} & \multicolumn{1}{c|}{\multirow{2}{*}{\begin{tabular}[c]{@{}c@{}}input\\ size\end{tabular}}} & \multirow{2}{*}{PCA} & \multirow{2}{*}{AutoEnc} \\
 &  & \multicolumn{1}{c|}{} &  &  \\ \midrule
\multirow{4}{*}{MNIST} & 26 & \multicolumn{1}{c|}{8} & $91.0\pm 12.7$ & $82.7\pm 15.2$ \\
 & 34 & \multicolumn{1}{c|}{16} & $97.0\pm 3.5$ & $83.5\pm 15.5$ \\
 & 44 & \multicolumn{1}{c|}{8} & $93.3\pm 13.2$ & $90.4\pm 13.4$ \\
 & 56 & \multicolumn{1}{c|}{16} & $93.0\pm 13.4$ & $95.5\pm 2.3$ \\ \midrule
\multirow{4}{*}{\begin{tabular}[c]{@{}c@{}}Fashion\\ MNIST\end{tabular}} & 26 & \multicolumn{1}{c|}{8} & $82.2\pm 16.6$ & $86.8\pm 12.7$ \\
 & 34 & \multicolumn{1}{c|}{16} & $78.8\pm 19.1$ & $79.0\pm 19.0$ \\
 & 44 & \multicolumn{1}{c|}{8} & $89.4\pm 3.9$ & $92.4\pm 2.8$ \\
 & 56 & \multicolumn{1}{c|}{16} & $91.9\pm 2.0$ & $93.6\pm 2.2$ \\ \bottomrule
\end{tabular}
\caption{\label{tab:cnn}Mean classification accuracy and one standard deviation obtained with classical CNN for classifying 0 and 1 in the MNIST and Fashion MNIST datasets. Each column is named with the pre-processing method (PCA or AutoEnc). These results directly compare to the second and third columns of Tabs.~\ref{tab:MNIST_ce} and \ref{tab:FMNIST_ce} denoted by Qubit and Dense.}
\end{table}

\begin{figure*}[ht]
    \centering
    \includegraphics[width=0.8\textwidth]{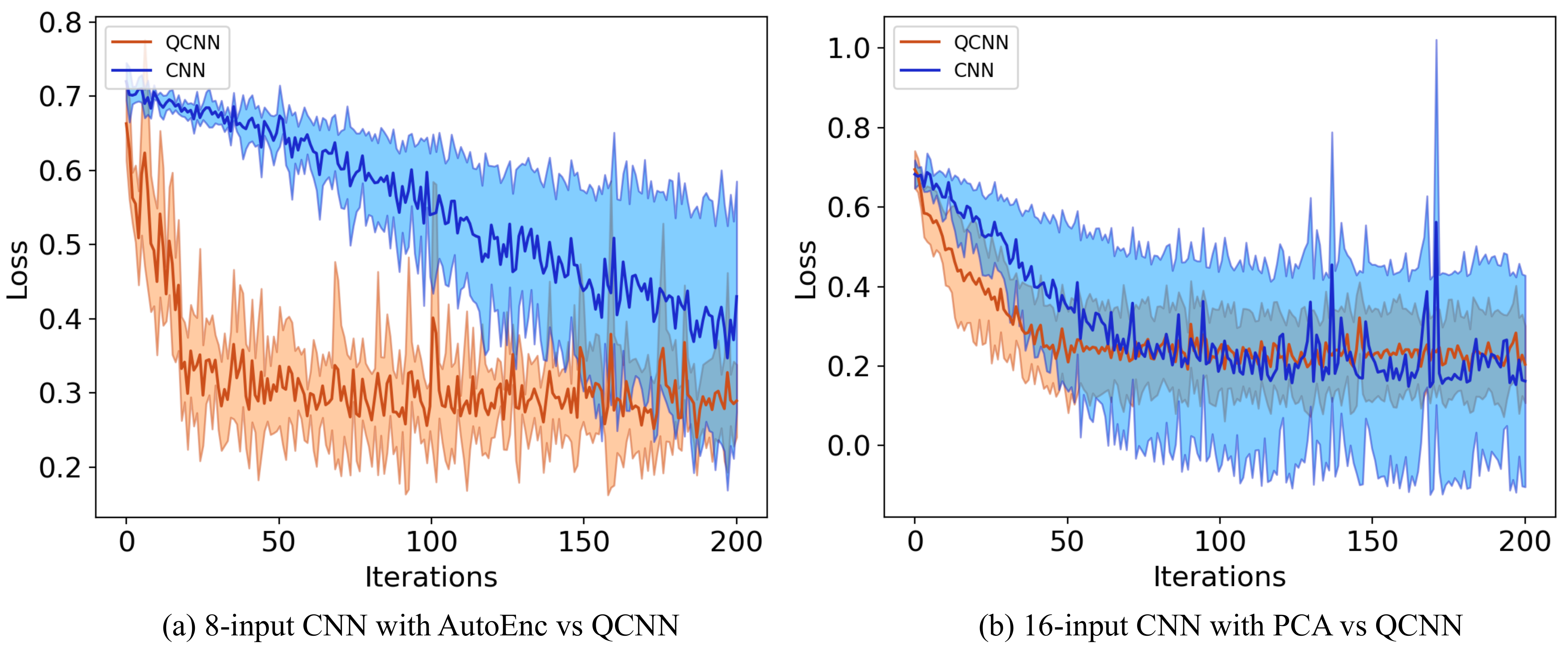}
    \caption{\label{fig:loss}Cross-entropy loss as a function of the number of training iterations. The QCNN models use circuit 9b as the ansatz. (a) The QCNN model with qubit encoding and autoencoding is compared to the CNN model with 8-inputs. (b) The QCNN model with dense encoding and PCA is compared to the CNN model with 16-inputs.}
\end{figure*}

In Fig.~\ref{fig:loss}, we present two representative examples of the cross-entropy loss as a function of the number of training iterations. For simplicity, we show such data for two cases in MNIST data classification: circuit 9b and qubit encoding with autoencoding, and circuit 9b and dense encoding with PCA. Considering the number of free parameters, these cases are comparable to the CNN models with 8 inputs with autoencoding and 16 inputs with PCA, respectively. Recall that the mean classification accuracy and one standard deviation in QCNN (CNN) is $96.6\pm 2.2$ ($90.4\pm 13.4$) for the first case, and $98.3\pm 0.5$ ($93.0\pm 13.4$) for the second case. Figure~\ref{fig:loss} shows that in both cases, the QCNN models are trained faster than the CNN models, while the advantage manifests more clearly in the first case. Furthermore, the standard deviations in the QCNN models are significantly smaller than that of the CNN models.

%
\section{Conclusion}
\label{sec:conclusion}

Fully parameterized quantum convolutional neural networks pave promising avenues for near-term applications of quantum machine learning and data science. This work presented an extensive benchmark of QCNN for solving classification problems on classical data, a fundamental task in pattern recognition. The QCNN algorithm can be tailored with many variables such as the structure of parameterized quantum circuits (i.e. ansatz) for convolutional filters and pooling operators, quantum data encoding methods, classical data pre-processing methods, cost functions and optimizers. To improve the utility of QCNN for classical data, we also introduced new data encoding schemes, namely hybrid direct encoding and hybrid angle encoding, with which the exchange between quantum circuit depth and width for state preparation can be configured. With diverse combinations of the aforementioned variables, we tested 8-qubit QCNN models for binary classification of MNIST and Fashion MNIST datasets by simulation with Pennylane. The QCNN models tested in this work operated with a small number of free parameters, ranging from 12 to 51. Despite the small number of free parameters, QCNN produced high classification accuracy for all instances, with the best case being close to $99\%$ for MNIST and $94\%$ for Fashion MNIST. We also compared QCNN results to CNN and observed that QCNN performed noticeably better than CNN given the similar training conditions for both benchmarking datasets. The comparison between QCNN and CNN is only valid for qubit and dense encoding cases in which the number of input qubits grows linearly with the dimension of the input data. With amplitude or hybrid encoding, the number of input qubits is substantially smaller than the dimension of the data, and hence there is no classical analogue. We speculate that the advantage of QCNN lies in the ability to exploit entanglement, which is a global effect, while CNN is only capable of capturing local correlations.

The QCNN architecture proposed in this work can be generalized for $L$-class classification through one-vs-one or one-vs-all strategies. It also remains an interesting future work to examine the construction of a multi-class classifier by leaving $\lceil\log_2(L)\rceil$ qubits for measurement in the output layer. Another interesting future work is to optimize the data encoding via training methods provided in Ref.~\cite{lloyd2020quantum}. However, since QCNN itself can be viewed as a feature reduction technique, it is not clear whether introducing another layer of the variational quantum circuit for data encoding would help until a thorough investigation is carried out. Understanding the underlying principle for the quantum advantage demonstrated in this work also remains to be done. One way to study this is by testing QCNN models with a set of data that does not exhibit local correlation but contains some global feature while analyzing the amount of entanglement created in the QCNN circuit. Since the circuit depth grows only logarithmically with the number of input qubits and the gate parameters are learned, the QCNN model is expected to be suitable for NISQ devices. However, the verification through real-world experiments and noisy simulations remains to be done. Furthermore, testing the classification performance as the QCNN models grow bigger remains an interesting future work. Finally, the application of the proposed QCNN algorithms for other real-world datasets such as those relevant to high-energy physics and medical diagnosis is of significant importance.

\section*{Acknowledgements}
This research is supported by the National Research Foundation of Korea (Grant No. 2019R1I1A1A01050161 and 2021M3H3A1038085) and Quantum Computing Development Program (Grant No. 2019M3E4A1080227). We thank Quantum Open Source Foundation as this work was initiated under the Quantum Computing Mentorship program.
\section*{Data availability}
The source code used in this study is available at \url{https://github.com/takh04/QCNN}.

\section*{Conflict of interest}

The authors declare that they have no conflict of interest.

\appendix
\section{Related works}
\label{sec:appA}
The term \textit{Quantum Convolutional Neural Network} (QCNN) appears in several places, but it refers to a number of different frameworks. Several proposals have been made in the past to reproduce classical CNN on a quantum circuit by imitating the basic arithmetic of the convolutional layer for a given filter~\cite{kerenidis2019quantum,li_quantum_2020,wei_quantum_2021}. Although these algorithms have the potential to achieve exponential speedups against the classical counterpart in the asymptotic limit, they require an efficient means to implement quantum random access memory (QRAM), expensive subroutines such as the linear combination of unitaries or quantum phase estimation with extra qubits, and they work only for specific types of quantum data embedding. Another branch of CNN-inspired QML algorithms focuses on implementing the convolutional filter as a parameterized quantum circuit, which can be stacked by inserting a classical pooling layer in between~\cite{liu2019hybrid,henderson_quanvolutional_2020,chen2020quantum}. Following the nomenclature provided in ~\cite{henderson_quanvolutional_2020}, we refer to this approach as \textit{quanvolutioanl} neural network to distinguish it from QCNN. The potential quantum advantage of using quanvolutional layers lies in the fact that quantum computers can access kernel functions in high-dimensional Hilbert spaces much more efficiently than classical computers. In quanvolutional NN, a challenge is to find a good structure for the parametric quantum circuit in which the number of qubits equals the size of the filter. This approach is also limited to qubit encoding since each layer requires a quantum embedding which has a non-negligible cost. Furthermore, stacking quanvolutional layers via pooling requires each parameterized quantum circuit to be measured multiple times for the measurement statistics.

Variational quantum circuits with the hierarchical structure consisting of $O(\log(n))$ layers do not exhibit the problem of ``barren plateau"~\cite{pesah2020absence}. In other words, the precision required in the measurement grows at most polynomially with the system size. This result guarantees the trainability of the fully parameterized QCNN models studied in this work when randomly initializing their parameters. Furthermore, numerical calculations in Ref.~\cite{pesah2020absence} show that the cost function gradient vanishes at a slower rate (with $n$, the number of initial qubits) when all unitary operators in the same layer are identical as in QCNN~\cite{cong_quantum_2019}. The hierarchical structure inspired by tensor network, without translational invariance, was first introduced in Ref.~\cite{grant_hierarchical_2018}. The hierarchical quantum circuit can be combined with a classical neural network as demonstrated in Ref.~\cite{HUANG202189}.

We note in passing that there exist several works proposing the quantum version of perceptron for binary classification~\cite{tacchino_artificial_2019,Mangini_2020,MONTEIRO2021698}. While our QCNN model differs from them as it implements the entire network as a parameterized quantum circuit, interesting future work is to investigate the alternative approach to construct a complex network of quantum artificial neurons developed in the previous works.

\section{Classical CNN}
\label{sec:AppB}
\begin{figure}[ht]
    \centering
    \includegraphics[width=0.95\columnwidth]{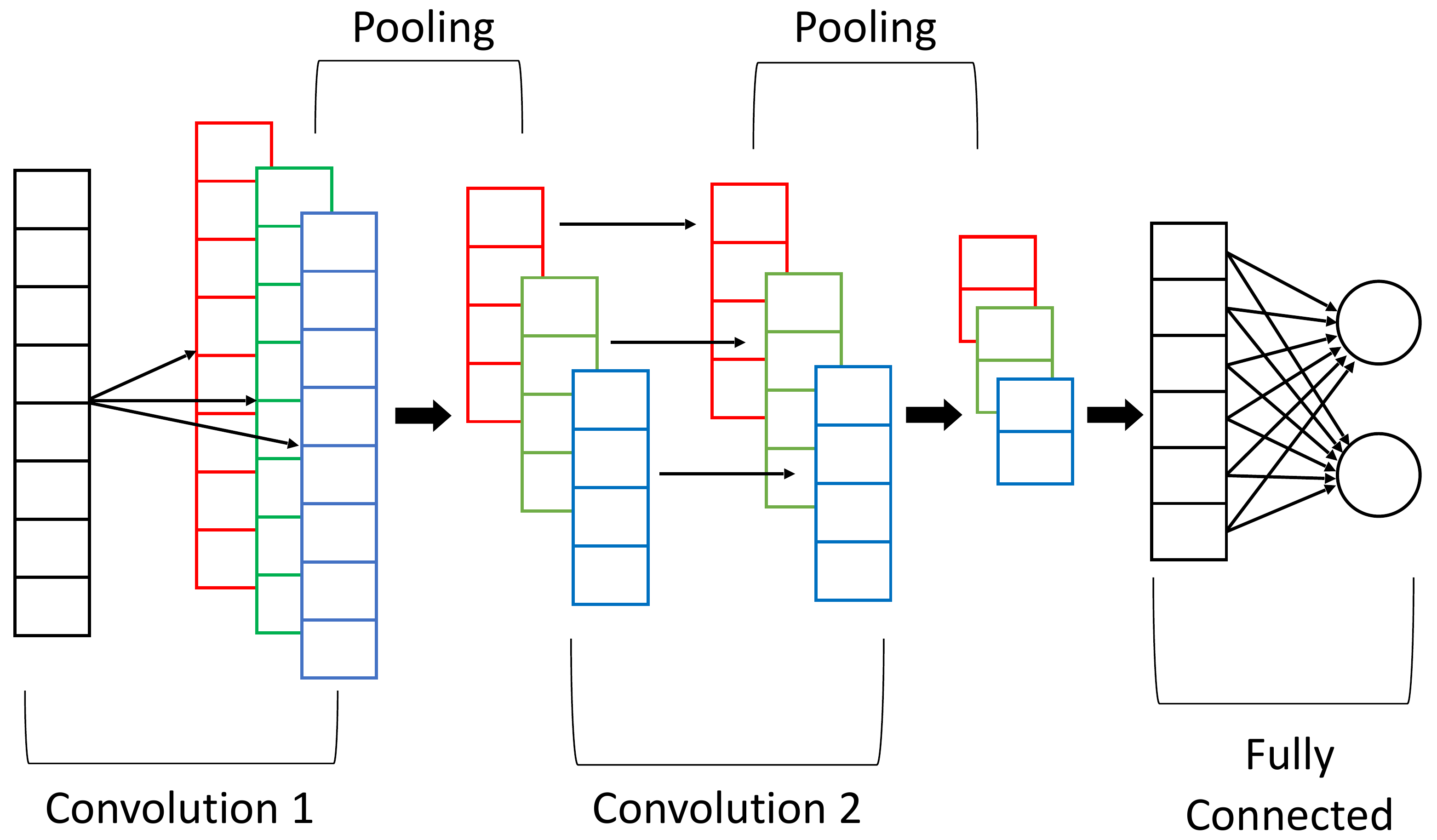}
    \caption{\label{fig:A1}A schematic of CNN used in this work for comparing to the classification performance of QCNN. To make the comparison as fair as possible, the number of free parameters are adjusted to be similar to that used in QCNN. This leads to starting with a small number of input nodes. While we used two CNN structures with 8 and 16 input nodes, the figure shows a CNN structure with 8 input nodes as an example.}
\end{figure}
In order to compare the classification accuracy of CNN and QCNN in fair conditions, we fixed hyperparameters used in the optimization step to be the same, which include iteration numbers, batch size, optimizer type, and its learning rates. In addition, we modified the structure of CNN in ways that its number of parameters subject to optimization is as close to that used in QCNN as possible. For example, since QCNN attains the best results with about 40 to 50 free parameters, we adjust the CNN structure accordingly. This led us to come up with two CNN, one with the input shape of (8, 1, 1) and another with the input shape of (16, 1, 1). In order to occupy the small number of input nodes for MNIST and Fashion MNIST classification, PCA and autoencoding are used for data pre-processing as done in QCNN. The CNNs go through convolutional and pooling stages twice, followed by a fully connected layer. The number of free parameters used in the CNN models is 26 or 44 for the case of 8 input nodes and 34 or 56 for the case of 16 input nodes. The training also mimics that of QCNN. For every iteration step, 25 data are randomly selected from the training data set, and trained via Adam optimizer with the learning rate of 0.01. We also fixed the number of iterations to be 200 as done in QCNN. The number of training (test) data are 12665 (2115) and 12000 (2000) for MNIST and fashion MNIST datasets, respectively. 

\section{QCNN simulation results for MSE loss}
\label{sec:AppC}
In Sec.~\ref{sec:simulation} of the main text, we presented the Pennylane simulation results of QCNN trained with the cross-entropy loss. When MSE is used as the cost function, similar results are obtained. We report classification results for MNIST and Fashion MNIST data attained from QCNN models that are trained with MSE in Tab.~\ref{tab:MNIST_mse} and \ref{tab:FMNIST_mse}.

\onecolumngrid

\begin{table}[h]
\centering
\begin{tabular}{@{}ccccccc@{}}
\toprule
\multicolumn{2}{c}{}        & \multicolumn{5}{c}{Classification Accuracy} \\ \midrule
\multirow{2}{*}{Ansatz} &
  \multirow{2}{*}{\begin{tabular}[c]{@{}c@{}}\# of \\ params\end{tabular}} &
  \multirow{2}{*}{Amplitude} &
  \multirow{2}{*}{Qubit} &
  \multirow{2}{*}{Dense} &
  \multirow{2}{*}{HDE} & 
  \multirow{2}{*}{HAE} \\
 &                        &           &           &          &  &         \\ \midrule
1 & \multicolumn{1}{c|}{12} &  $92.4\pm 3.1$ &  \begin{tabular}[c]{@{}c@{}}$96.1\pm 3.3$ \\ $86.1\pm 8.8$\end{tabular}         &   \begin{tabular}[c]{@{}c@{}} $91.4\pm 8.6$    \\ $88.9\pm 7.1$\end{tabular}        &   \begin{tabular}[c]{@{}c@{}} $60.3 \pm 3.0$    \\ $83.4 \pm 4.0 $ \end{tabular}   &   \begin{tabular}[c]{@{}c@{}} $97.8 \pm 1.2$   \\ $77.9 \pm 6.1$  \end{tabular}   \\
2 & \multicolumn{1}{c|}{12} & $88.2\pm 7.1$ & \begin{tabular}[c]{@{}c@{}} $86.4\pm 7.1$ \\ $84.2\pm 10.0$\end{tabular} & \begin{tabular}[c]{@{}c@{}} $91.4 \pm 0.3$    \\ $88.9 \pm 1.4$\end{tabular}     & \begin{tabular}[c]{@{}c@{}} $54.4 \pm 4.6$    \\ $87.5 \pm 8.6$\end{tabular} & \begin{tabular}[c]{@{}c@{}} $93.0 \pm 2.7$    \\ $78.7 \pm 6.6$\end{tabular}        \\
3 & \multicolumn{1}{c|}{18} & $94.1\pm 1.7$ & \begin{tabular}[c]{@{}c@{}} $98.0\pm 1.4$ \\ $91.6\pm 8.9$\end{tabular} &   \begin{tabular}[c]{@{}c@{}} $98.3\pm 0.1$ \\ $95.3\pm 3.2$\end{tabular}        & \begin{tabular}[c]{@{}c@{}} $69.8 \pm 6.2$    \\ $94.2 \pm 3.8$\end{tabular}     & \begin{tabular}[c]{@{}c@{}} $98.3 \pm 0.3$    \\ $83.5 \pm 3.9$\end{tabular}   \\
4 & \multicolumn{1}{c|}{24} & $90.1\pm 2.0$ & \begin{tabular}[c]{@{}c@{}} $98.2\pm 0.1  $ \\ $88.8\pm 6.3$\end{tabular} &  \begin{tabular}[c]{@{}c@{}} $84.9\pm 2.5$ \\ $85.8 \pm 6.3$\end{tabular}         & \begin{tabular}[c]{@{}c@{}} $63.2 \pm 3.7$    \\ $94.8 \pm 1.1$\end{tabular}     & \begin{tabular}[c]{@{}c@{}} $98.0 \pm 0.2$    \\ $85.5 \pm 2.0$\end{tabular}   \\
5 & \multicolumn{1}{c|}{24} & $91.9\pm 1.7$ & \begin{tabular}[c]{@{}c@{}} $98.1\pm 0.1$ \\ $92.7\pm 3.5$\end{tabular} &  \begin{tabular}[c]{@{}c@{}} $98.3\pm 0.1$ \\ $94.0\pm 2.0$\end{tabular}         &  \begin{tabular}[c]{@{}c@{}} $65.7\pm 2.3$ \\ $91.8\pm 3.5$\end{tabular}    &  \begin{tabular}[c]{@{}c@{}} $98.0\pm 0.1$ \\ $83.0\pm 4.5$\end{tabular}    \\
6 & \multicolumn{1}{c|}{24} & $96.2\pm 2.0$ & \begin{tabular}[c]{@{}c@{}} $98.1\pm 0.1$ \\ $94.4\pm 3.9$\end{tabular} &   \begin{tabular}[c]{@{}c@{}} $97.8\pm 0.2$ \\ $92.5\pm 1.7$\end{tabular}          &  \begin{tabular}[c]{@{}c@{}} $74.4\pm 3.9$ \\ $94.8\pm 2.8$\end{tabular}    &  \begin{tabular}[c]{@{}c@{}} $98.3\pm 0.2$ \\ $\bold{86.6}\pm 4.2$\end{tabular}  \\  
7 & \multicolumn{1}{c|}{36} & $93.2\pm 4.3$ & \begin{tabular}[c]{@{}c@{}} $98.4\pm 0.1$ \\ $95.2\pm 4.5$\end{tabular} &   \begin{tabular}[c]{@{}c@{}} $98.0\pm 0.3$ \\ $92.4\pm 3.0$\end{tabular}      &   \begin{tabular}[c]{@{}c@{}} $68.9\pm 3.8$ \\ $94.7\pm 4.1$\end{tabular}   &  \begin{tabular}[c]{@{}c@{}} $98.1\pm 0.2$ \\ $80.2\pm 3.5$\end{tabular}  \\
8 & \multicolumn{1}{c|}{36} & $95.2\pm 1.4$ & \begin{tabular}[c]{@{}c@{}} $\bold{98.5}\pm 0.1$ \\ $\bold{97.0}\pm 2.8$\end{tabular} &   \begin{tabular}[c]{@{}c@{}} $98.3\pm 0.1$ \\ $93.4\pm 2.0$\end{tabular}         & \begin{tabular}[c]{@{}c@{}} $68.5\pm 3.7$ \\ $94.0\pm 2.8$\end{tabular}     &   \begin{tabular}[c]{@{}c@{}} $98.2\pm 0.1$ \\ $86.1\pm 4.5$\end{tabular} \\
9a & \multicolumn{1}{c|}{51} & $97.0\pm 1.0$ & \begin{tabular}[c]{@{}c@{}} $98.2\pm 1.0$ \\  $96.2\pm 1.4$\end{tabular} &    \begin{tabular}[c]{@{}c@{}} $98.4 \pm 0.1$ \\  $\bold{96.6} \pm 1.7$\end{tabular}        &   \begin{tabular}[c]{@{}c@{}} $\bold{77.4}\pm 1.8$ \\ $\textbf{97.2}\pm 1.8$\end{tabular}   &   \begin{tabular}[c]{@{}c@{}} $\textbf{98.5}\pm 0.2$ \\ $85.0\pm 5.8$\end{tabular} \\  
9b & \multicolumn{1}{c|}{45} & $\bold{98.4}\pm 0.1$ & \begin{tabular}[c]{@{}c@{}} $\bold{98.5}\pm 0.3$ \\ $\bold{97.0}\pm 2.0$\end{tabular} &   \begin{tabular}[c]{@{}c@{}} $\bold{98.5}\pm 0.1$ \\ $95.7\pm 1.7$\end{tabular}       &   \begin{tabular}[c]{@{}c@{}} $75.9\pm 0.9$ \\ $96.7\pm 1.7$\end{tabular}     & \begin{tabular}[c]{@{}c@{}} $98.3\pm 0.3$ \\ $85.9\pm 3.8$\end{tabular} \\\bottomrule
\end{tabular}
\caption{\label{tab:MNIST_mse}Mean accuracy and one standard deviation of the classification for 0 and 1 in the MNIST dataset when the QCNN model is trained with MSE. The mean and the standard deviation are obtained from five repetitions with random initialization of parameters. The first column shows the ansatz label. The second column shows the total number of parameters that are subject to optimization. For qubit, dense and hybrid encoding, two rows indicate the values obtained with different classical data pre-processing, namely PCA and autoencoding, respectively. The best result under each quantum data encoding method is written in bold.}
\end{table}

\begin{table}[h]
\centering
\begin{tabular}{@{}ccccccc@{}}
\toprule
\multicolumn{2}{c}{}        & \multicolumn{5}{c}{Classification Accuracy} \\ \midrule
\multirow{2}{*}{Ansatz} &
  \multirow{2}{*}{\begin{tabular}[c]{@{}c@{}}$\#$ of \\ params\end{tabular}} &
  \multirow{2}{*}{Amplitude} &
  \multirow{2}{*}{Qubit} &
  \multirow{2}{*}{Dense} &
  \multirow{2}{*}{HDE} & \multirow{2}{*}{HAE}\\
 &  &                       &           &           &          &          \\ \midrule
 1  & \multicolumn{1}{c|}{12} & $88.1\pm 3.0$ & \begin{tabular}[c]{@{}c@{}} $79.6\pm 10.9$ \\ $80.0\pm 2.4$\end{tabular} & \begin{tabular}[c]{@{}c@{}} $81.4 \pm 5.5$    \\ $77.8 \pm 8.4$\end{tabular}     &  \begin{tabular}[c]{@{}c@{}} $58.6 \pm 2.4$    \\ $89.7 \pm 1.8$\end{tabular}    &   \begin{tabular}[c]{@{}c@{}} $84.2 \pm 1.3$    \\ $83.1 \pm 4.3$\end{tabular}   \\
2 & \multicolumn{1}{c|}{12} &  $87.8\pm 3.0$ &  \begin{tabular}[c]{@{}c@{}}$80.0\pm 8.4$ \\ $70.0\pm 10.0$\end{tabular}         &   \begin{tabular}[c]{@{}c@{}} $78.0 \pm 5.2$    \\ $78.5 \pm 8.3$\end{tabular}        &   \begin{tabular}[c]{@{}c@{}} $54.1 \pm 3.5$    \\ $88.7 \pm 2.8$\end{tabular}   &   \begin{tabular}[c]{@{}c@{}} $78.8 \pm 4.5$    \\ $81.8 \pm 5.1$\end{tabular}   \\
3 & \multicolumn{1}{c|}{18} & $87.0\pm 3.0$ & \begin{tabular}[c]{@{}c@{}} $87.6\pm 3.6$ \\ $85.1\pm 5.7$\end{tabular} &   \begin{tabular}[c]{@{}c@{}} $92.7\pm 2.4$ \\ $84.7\pm 7.2$\end{tabular}        &   \begin{tabular}[c]{@{}c@{}} $61.2\pm 2.9$ \\ $90.1\pm 1.1$\end{tabular}   &  \begin{tabular}[c]{@{}c@{}} $86.0\pm 2.6$ \\ $86.2\pm 3.3$\end{tabular}  \\
4 & \multicolumn{1}{c|}{24} & $89.7\pm 1.3$ & \begin{tabular}[c]{@{}c@{}} $85.4\pm 2.9$ \\ $81.2\pm 2.2$\end{tabular} &  \begin{tabular}[c]{@{}c@{}} $90.7\pm 1.5$ \\ $84.2\pm 5.6$\end{tabular}         &  \begin{tabular}[c]{@{}c@{}} $62.5\pm 2.0$ \\ $88.4\pm 4.9$\end{tabular}    &  \begin{tabular}[c]{@{}c@{}} $84.0\pm 1.7$ \\ $88.8\pm 3.6$\end{tabular}  \\
5 & \multicolumn{1}{c|}{24} & $90.7\pm 1.2$ & \begin{tabular}[c]{@{}c@{}} $83.8\pm 4.4$ \\ $81.1\pm 1.4$\end{tabular} &  \begin{tabular}[c]{@{}c@{}} $88.8\pm 3.0$ \\ $86.2\pm 3.3$\end{tabular}         &  \begin{tabular}[c]{@{}c@{}} $60.4\pm 0.9$ \\ $90.7\pm 1.9$\end{tabular}    &  \begin{tabular}[c]{@{}c@{}} $84.7\pm 3.5$ \\ $84.9\pm 4.8$\end{tabular}  \\ 
6 & \multicolumn{1}{c|}{24} & $88.8\pm 3.0$ & \begin{tabular}[c]{@{}c@{}} $86.6\pm 2.4$ \\ $86.1\pm 3.0$\end{tabular} &   \begin{tabular}[c]{@{}c@{}} $89.4\pm 4.4$ \\ $84.6\pm 5.3$\end{tabular}          &     \begin{tabular}[c]{@{}c@{}} $63.9\pm 2.2$ \\ $91.7\pm 1.8$\end{tabular} &  \begin{tabular}[c]{@{}c@{}} $84.4\pm 1.0$ \\ $85.8\pm 4.0$\end{tabular}  \\
7 & \multicolumn{1}{c|}{36} & $90.0\pm 1.2$ & \begin{tabular}[c]{@{}c@{}} $85.4\pm 3.2$ \\ $86.4\pm 2.7$\end{tabular} &   \begin{tabular}[c]{@{}c@{}} $92.5\pm 3.6$ \\ $87.7\pm 6.1$\end{tabular}      &   \begin{tabular}[c]{@{}c@{}} $64.4\pm 1.8$ \\ $91.9\pm 1.8$\end{tabular}    &  \begin{tabular}[c]{@{}c@{}} $84.6\pm 0.8$ \\ $89.0\pm 3.2$\end{tabular}   \\
8 & \multicolumn{1}{c|}{36} & $89.7\pm 2.8$ & \begin{tabular}[c]{@{}c@{}} $82.3\pm 2.1$ \\ $85.8\pm 2.2$\end{tabular} &   \begin{tabular}[c]{@{}c@{}} $90.0\pm 2.3$ \\ $90.1\pm 3.6$\end{tabular}         &  \begin{tabular}[c]{@{}c@{}} $64.9\pm 2.5$ \\ $\bold{92.9}\pm 0.7$\end{tabular}    &  \begin{tabular}[c]{@{}c@{}} $86.5\pm 1.9$ \\ $86.3\pm 7.2$\end{tabular}  \\
9a & \multicolumn{1}{c|}{51} & $90.8\pm 1.2$ & \begin{tabular}[c]{@{}c@{}} $\bold{89.9}\pm 1.9$ \\  $\bold{92.7}\pm 0.4$\end{tabular} &    \begin{tabular}[c]{@{}c@{}} ${88.8} \pm 1.9$ \\  $\bold{93.3} \pm 1.1$\end{tabular}        &  \begin{tabular}[c]{@{}c@{}} $\bold{66.8} \pm 1.7$ \\ $92.8\pm 1.3$\end{tabular}    & \begin{tabular}[c]{@{}c@{}} $88.4\pm 1.4$ \\ $\textbf{90.8}\pm 2.1$\end{tabular}   \\ 
9b & \multicolumn{1}{c|}{45} & $\bold{91.0}\pm 1.1$ & \begin{tabular}[c]{@{}c@{}} ${89.4}\pm 2.5$ \\ $89.1\pm 2.0$\end{tabular} &   \begin{tabular}[c]{@{}c@{}} $\bold{93.0}\pm 1.1$ \\ ${90.4}\pm 3.1$\end{tabular}       &  \begin{tabular}[c]{@{}c@{}} $65.6\pm 2.6$ \\ $\textbf{92.9}\pm 1.6$\end{tabular}    &   \begin{tabular}[c]{@{}c@{}} $\bold{89.3}\pm 1.5$ \\ $89.2\pm 4.6$\end{tabular}   \\  \bottomrule
\end{tabular}
\caption{\label{tab:FMNIST_mse}Mean accuracy and one standard deviation of the classification for 0 (t-shirt/top) and 1 (trouser) in the Fashion MNIST dataset when the QCNN model is trained with MSE. The mean and the standard deviation are obtained from five repetitions with random initialization of parameters. The first column shows the ansatz label. The second column shows the total number of parameters that are subject to optimization. For qubit, dense and hybrid encoding, two rows indicate the values obtained with different classical data pre-processing, namely PCA and autoencoding, respectively. The best result under each quantum data encoding method is written in bold.}
\end{table}

\clearpage
\twocolumngrid

\section{Classification with Hierarchical Quantum Classifier}
\label{sec:HQC}
The hierarchical structure inspired by tensor network named as hierarchical quantum classifier (HQC) was first introduced in Ref.~\cite{grant_hierarchical_2018}. The HQC therein does not enforce translational invariance, and hence the number of free parameters subject to optimization grows as $O(n)$ for a quantum circuit with $n$ input qubits. Although the simulation presented in the main manuscript aims to benchmark the classification performance of the QML model in which the number of parameters grows as $O(\log(n))$, we also report the simulation results of HQC with the tensor tree network (TTN) structure~\cite{grant_hierarchical_2018} in this supplementary section for interested readers. The TTN classifier does not employ parameterized quantum gates for pooling. Thus for certain ansatz, the number of parameters differs from that of QCNN models. For example, although the convolutional circuit 2 in Fig.~\ref{fig:convolution} has two free parameters, only one of them is effective since one of the qubits is traced out as soon as the parameterized gate is applied. For brevity, here we only report the results obtained with the cross-entropy loss but similar results can be obtained with MSE. As can be seen from Tab.~\ref{tab:HQC_MNIST} and Tab.~\ref{tab:HQC_FMNIST}, the number of effective parameters (i.e. the second column) grows faster that that of QCNN models. An interesting observation is that there is no clear trend as the number of parameters is increased beyond 42, which is close to the maximum number of parameters used in QCNN. In other words, there is no clear motivation to increase the number of free parameters beyond 42 or so when seeking to improve the classification performance. Studying overfitting under the growth of the number of parameters remains an interesting open problem.

\onecolumngrid

\vspace{1em}
\begin{table}[h]
\centering
\begin{tabular}{@{}ccccccc@{}}
\toprule
\multicolumn{2}{c}{}        & \multicolumn{5}{c}{Classification Accuracy} \\ \midrule
\multirow{2}{*}{Ansatz} &
  \multirow{2}{*}{\begin{tabular}[c]{@{}c@{}}$\#$ of \\ params\end{tabular}} &
  \multirow{2}{*}{Amplitude} &
  \multirow{2}{*}{Qubit} &
  \multirow{2}{*}{Dense} &
  \multirow{2}{*}{HDE} & \multirow{2}{*}{HAE}\\
 &  &                       &           &           &          &          \\ \midrule
 1  & \multicolumn{1}{c|}{14} & $96.5\pm 0.4$ & \begin{tabular}[c]{@{}c@{}} $94.7\pm 2.2$ \\ $91.8\pm 1.2$\end{tabular} & \begin{tabular}[c]{@{}c@{}} $50.5\pm 3.5$    \\ $82.1\pm 6.5$\end{tabular}     &  \begin{tabular}[c]{@{}c@{}} $69.9\pm 1.4$    \\ $94.2\pm 3.0$\end{tabular}    &   \begin{tabular}[c]{@{}c@{}} $\bold{98.5}\pm 0.1$    \\ $81.6\pm 2.3$\end{tabular}   \\
2 & \multicolumn{1}{c|}{7} & $55.6\pm 1.7$ & \begin{tabular}[c]{@{}c@{}} $57.3\pm 3.2$ \\ $64.5\pm 15.9$\end{tabular} &  \begin{tabular}[c]{@{}c@{}} $52.7\pm 1.2$ \\ $74.0\pm 3.4$\end{tabular}         &  \begin{tabular}[c]{@{}c@{}} $53.8\pm 0.2$ \\ $76.4\pm 8.9$\end{tabular}    &  \begin{tabular}[c]{@{}c@{}} $60.3\pm 0.2$ \\ $56.1\pm 6.9$\end{tabular}   \\
3 & \multicolumn{1}{c|}{28} & $98.7\pm 0.2$ & \begin{tabular}[c]{@{}c@{}} $98.2\pm 0.5$ \\ $97.7\pm 1.2$\end{tabular} &   \begin{tabular}[c]{@{}c@{}} $97.8\pm 0.2$ \\ $95.7\pm 1.8$\end{tabular}      &   \begin{tabular}[c]{@{}c@{}} $81.2\pm 0.9$ \\ $97.8\pm 1.4$\end{tabular}    &  \begin{tabular}[c]{@{}c@{}} $98.3\pm 0.1$ \\ $\bold{88.7}\pm 4.3$\end{tabular}  \\

4 & \multicolumn{1}{c|}{42} & $95.6\pm 2.2$ & \begin{tabular}[c]{@{}c@{}} $87.4\pm 15.0$ \\ $95.9\pm 1.1$\end{tabular} &  \begin{tabular}[c]{@{}c@{}} $84.2\pm 15.5$ \\ $95.7\pm 1.4$\end{tabular}         &  \begin{tabular}[c]{@{}c@{}} $77.0\pm 2.2$ \\ $89.5\pm 8.0$\end{tabular}    &  \begin{tabular}[c]{@{}c@{}} $90.0\pm 10.3$ \\ $81.5\pm 3.0$\end{tabular}\\

5 & \multicolumn{1}{c|}{42} & $97.8\pm 0.5$ & \begin{tabular}[c]{@{}c@{}} $96.7\pm 2.0$ \\ $96.0\pm 3.1$\end{tabular} &   \begin{tabular}[c]{@{}c@{}} $97.0\pm 1.1$ \\ $96.5\pm 1.0$\end{tabular}          &     \begin{tabular}[c]{@{}c@{}} $77.1\pm 0.8$ \\ $94.1\pm 3.4$\end{tabular} &  \begin{tabular}[c]{@{}c@{}} $98.2\pm 0.2$ \\ $88.1\pm 2.1$\end{tabular}  \\ 
6 & \multicolumn{1}{c|}{35} & $98.6\pm 0.2$ & \begin{tabular}[c]{@{}c@{}} $98.0\pm 0.8$ \\ $98.1\pm 1.5$\end{tabular} &   \begin{tabular}[c]{@{}c@{}} $97.8\pm 0.1$ \\ $95.7\pm 3.1$\end{tabular}         &  \begin{tabular}[c]{@{}c@{}} $\bold{81.6}\pm 0.4$ \\ $98.2\pm 0.9$\end{tabular}    &  \begin{tabular}[c]{@{}c@{}} $98.3\pm 0.3$ \\ $90.2\pm 3.0$\end{tabular} \\
7 & \multicolumn{1}{c|}{56} &  $89.9\pm 13.0$ &  \begin{tabular}[c]{@{}c@{}}$88.7\pm 14.8$ \\ $92.8\pm 1.5$\end{tabular}         &   \begin{tabular}[c]{@{}c@{}} $63.0\pm 5.1$    \\ $94.9\pm 1.6$\end{tabular}        &   \begin{tabular}[c]{@{}c@{}} $72.4\pm 0.4$    \\ $90.2\pm 7.1$\end{tabular}   &   \begin{tabular}[c]{@{}c@{}} $93.0\pm 10.8$    \\ $85.2\pm 1.9$\end{tabular}\\

8 & \multicolumn{1}{c|}{56} & $98.6\pm 0.2$ & \begin{tabular}[c]{@{}c@{}} $95.8\pm 2.5$ \\ $\bold{98.3}\pm 1.0$\end{tabular} &   \begin{tabular}[c]{@{}c@{}} $97.5\pm 0.6$ \\ $\bold{97.9}\pm 0.9$\end{tabular}        &   \begin{tabular}[c]{@{}c@{}} $77.8\pm 1.8$ \\ $96.7\pm 1.1$\end{tabular}   &  \begin{tabular}[c]{@{}c@{}} $92.2\pm 12.2$ \\ $85.7\pm 7.1$\end{tabular} \\

9 & \multicolumn{1}{c|}{84} & $\bold{98.9}\pm 0.1$ & \begin{tabular}[c]{@{}c@{}} $\bold{98.6}\pm 0.2$ \\  $97.6\pm 0.9$\end{tabular} &    \begin{tabular}[c]{@{}c@{}} $\bold{98.6}\pm 0.2$ \\  $96.9\pm 0.6$\end{tabular}        &  \begin{tabular}[c]{@{}c@{}} $81.0\pm 5.4$ \\ $\bold{98.6}\pm 0.2$\end{tabular}    & \begin{tabular}[c]{@{}c@{}} $98.3\pm 0.0$ \\ $82.2\pm 5.5$\end{tabular}   \\  \bottomrule
\end{tabular}
\caption{\label{tab:HQC_MNIST}Mean accuracy and one standard deviation of the classification for 0 and 1 in the MNIST dataset when the HQC model is trained with cross-entropy loss. The mean and the standard deviation are obtained from five repetitions with random initialization of parameters. The first column shows the ansatz label. The second column shows the total number of parameters that are subject to optimization. For qubit, dense and hybrid encoding, two rows indicate the values obtained with different classical data pre-processing, namely PCA and autoencoding, respectively. The best result under each quantum data encoding method is written in bold.}
\end{table}

\begin{table}[h]
\centering
\begin{tabular}{@{}ccccccc@{}}
\toprule
\multicolumn{2}{c}{}        & \multicolumn{5}{c}{Classification Accuracy} \\ \midrule
\multirow{2}{*}{Ansatz} &
  \multirow{2}{*}{\begin{tabular}[c]{@{}c@{}}$\#$ of \\ params\end{tabular}} &
  \multirow{2}{*}{Amplitude} &
  \multirow{2}{*}{Qubit} &
  \multirow{2}{*}{Dense} &
  \multirow{2}{*}{HDE} & \multirow{2}{*}{HAE}\\
 &  &                       &           &           &          &          \\ \midrule
 1  & \multicolumn{1}{c|}{14} & $90.1\pm 2.27$ & \begin{tabular}[c]{@{}c@{}} $89.7\pm 2.3$ \\ $89.0\pm 3.6$\end{tabular} & \begin{tabular}[c]{@{}c@{}} $61.8\pm 12.8$    \\ $76.2\pm 12.2$\end{tabular}     &  \begin{tabular}[c]{@{}c@{}} $87.7\pm 1.9$    \\ $85.6\pm 5.5$\end{tabular}    &   \begin{tabular}[c]{@{}c@{}} $61.2\pm 1.1$    \\ $83.4\pm 18.8$\end{tabular}   \\
2 & \multicolumn{1}{c|}{7} & $77.7\pm 5.7$ & \begin{tabular}[c]{@{}c@{}} $66.5\pm 0.0$ \\ $63.5\pm 9.4$\end{tabular} &  \begin{tabular}[c]{@{}c@{}} $52.6\pm 13.3$ \\ $55.9\pm 6.8$\end{tabular}         &  \begin{tabular}[c]{@{}c@{}} $60.3\pm 0.0$ \\ $64.7\pm 10.1$\end{tabular}    &  \begin{tabular}[c]{@{}c@{}} $49.5\pm 0.3$ \\ $69.0\pm 13.4$\end{tabular}   \\
3 & \multicolumn{1}{c|}{28} & $89.5\pm 2.0$ & \begin{tabular}[c]{@{}c@{}} $90.4\pm 2.6$ \\ $93.3\pm 1.0$\end{tabular} &   \begin{tabular}[c]{@{}c@{}} $85.0\pm 2.3$ \\ $93.7\pm 1.7$\end{tabular}      &   \begin{tabular}[c]{@{}c@{}} $89.4\pm 1.7$ \\ $91.0\pm 1.8$\end{tabular}    &  \begin{tabular}[c]{@{}c@{}} $68.7\pm 3.1$ \\ $\bold{93.8}\pm 1.7$\end{tabular}  \\

4 & \multicolumn{1}{c|}{42} & $91.4\pm 1.2$ & \begin{tabular}[c]{@{}c@{}} $75.0\pm 22.7$ \\ $92.2\pm 1.1$\end{tabular} &  \begin{tabular}[c]{@{}c@{}} $\bold{91.6}\pm 0.5$ \\ $94.4\pm 0.4$\end{tabular}         &  \begin{tabular}[c]{@{}c@{}} $88.0\pm 1.3$ \\ $82.0\pm 5.0$\end{tabular}    &  \begin{tabular}[c]{@{}c@{}} $63.8\pm 2.5$ \\ $84.2\pm 19.1$\end{tabular}\\

5 & \multicolumn{1}{c|}{42} & $90.1\pm 2.0$ & \begin{tabular}[c]{@{}c@{}} $83.7\pm 18.4$ \\ $93.0\pm 1.4$\end{tabular} &   \begin{tabular}[c]{@{}c@{}} $92.0\pm 0.9$ \\ $91.0\pm 2.1$\end{tabular}          &     \begin{tabular}[c]{@{}c@{}} $89.2\pm 2.5$ \\ $86.2\pm 3.4$\end{tabular} &  \begin{tabular}[c]{@{}c@{}} $66.3\pm 1.5$ \\ $75.9\pm 23.7$\end{tabular}  \\ 
6 & \multicolumn{1}{c|}{35} & $\bold{92.5}\pm 1.1$ & \begin{tabular}[c]{@{}c@{}} $88.8\pm 1.8$ \\ $\bold{94.2}\pm 1.1$\end{tabular} &   \begin{tabular}[c]{@{}c@{}} $86.6\pm 1.8$ \\ $92.3\pm 1.5$\end{tabular}         &  \begin{tabular}[c]{@{}c@{}} $\bold{89.8}\pm 1.9$ \\ $\bold{91.4}\pm 3.4$\end{tabular}    &  \begin{tabular}[c]{@{}c@{}} $68.9\pm 2.9$ \\ $67.4\pm 23.8$\end{tabular} \\
7 & \multicolumn{1}{c|}{56} &  $90.9\pm 0.6$ &  \begin{tabular}[c]{@{}c@{}}$85.3\pm 14.0$ \\ $91.6\pm 3.4$\end{tabular}         &   \begin{tabular}[c]{@{}c@{}} $76.4\pm 2.8$    \\ $92.3\pm 2.5$\end{tabular}        &   \begin{tabular}[c]{@{}c@{}} $81.4\pm 6.9$    \\ $81.4\pm 8.7$\end{tabular}   &   \begin{tabular}[c]{@{}c@{}} $62.9\pm 1.8$    \\ $67.1\pm 23.4$\end{tabular}\\

8 & \multicolumn{1}{c|}{56} & $90.7\pm 2.4$ & \begin{tabular}[c]{@{}c@{}} $\bold{90.8}\pm 0.8$ \\ $94.1\pm 1.1$\end{tabular} &   \begin{tabular}[c]{@{}c@{}} $91.1\pm 0.4$ \\ $92.8\pm 1.0$\end{tabular}        &   \begin{tabular}[c]{@{}c@{}} $89.5\pm 1.7$ \\ $88.7\pm 3.2$\end{tabular}   &  \begin{tabular}[c]{@{}c@{}} $67.6\pm 1.9$ \\ $84.9\pm 19.5$\end{tabular} \\

9 & \multicolumn{1}{c|}{84} & $89.9\pm 1.9$ & \begin{tabular}[c]{@{}c@{}} $92.0\pm 1.6$ \\  $93.8\pm 0.7$\end{tabular} &    \begin{tabular}[c]{@{}c@{}} $91.3\pm 2.2$ \\  $\bold{94.5}\pm 0.8$\end{tabular}        &  \begin{tabular}[c]{@{}c@{}} $89.6\pm 1.0$ \\ $91.0\pm 1.8$\end{tabular}    & \begin{tabular}[c]{@{}c@{}} $\bold{69.0}\pm 1.7$ \\ $85.4\pm 19.8$\end{tabular}   \\  \bottomrule
\end{tabular}
\caption{\label{tab:HQC_FMNIST}Mean accuracy and one standard deviation of the classification for 0 (t-shirt/top) and 1 (trouser) in the Fashion MNIST dataset when the HQC model is trained with cross-entropy loss. The mean and the standard deviation are obtained from five repetitions with random initialization of parameters. The first column shows the ansatz label. The second column shows the total number of parameters that are subject to optimization. For qubit, dense and hybrid encoding, two rows indicate the values obtained with different classical data pre-processing, namely PCA and autoencoding, respectively. The best result under each quantum data encoding method is written in bold.}
\end{table}

\clearpage
\twocolumngrid

%

%
\end{document}